\documentclass[twocolumn]{aastex63}
\usepackage{booktabs}
\usepackage[version=4]{mhchem}
\usepackage{cancel}
\usepackage{amsmath,amsfonts,amssymb}
\received{February 23, 2021}
\revised{July 13, 2021}
\def \tex{\ensuremath{T_{{\rm ex}}}}
\def \ntot{\ensuremath{N_{\rm {tot}}}}
\def \dix#1{\ensuremath{{\,\rm 10^{#1}}}}
\def \tdix#1{\ensuremath{{\,\times 10^{#1}}}}
\def \ccc{\ensuremath{{\,\rm cm^{-3}}}}
\def \cc{\ensuremath{{\,\rm cm^{-2}}}}
\def \s{\ensuremath{{\,\rm s^{-1}}}}
\shorttitle{MAPS IX: Disk S-chemistry}
\shortauthors{Le Gal et al.}
\graphicspath{{./}{figures/}}
\begin{document}

\title{Molecules with ALMA at Planet-forming Scales (MAPS) XII: \\Inferring the C/O and S/H ratios in Protoplanetary Disks with Sulfur Molecules}
\correspondingauthor{Romane Le Gal}
\email{Romane.Le-Gal@univ-grenoble-alpes.fr}
\author[0000-0003-1837-3772]{Romane Le Gal}
\altaffiliation{CNES Fellowship Program Fellow}
\affiliation{Center for Astrophysics \textbar\, Harvard \& Smithsonian, 60 Garden St., Cambridge, MA 02138, USA}
\affiliation{IRAP, Université de Toulouse, CNRS, CNES, UT3, F-31000 Toulouse, France}
\affiliation{Universit\'e Grenoble Alpes, CNRS, IPAG, F-38000 Grenoble, France}
\affiliation{IRAM, 300 rue de la Piscine, F-38406 Saint-Martin d'Hères, France}

\author[0000-0001-8798-1347]{Karin I. \"Oberg}
\affiliation{Center for Astrophysics \textbar\, Harvard \& Smithsonian, 60 Garden St., Cambridge, MA 02138, USA}

\author[0000-0003-1534-5186]{Richard Teague}
\affiliation{Center for Astrophysics \textbar\, Harvard \& Smithsonian, 60 Garden St., Cambridge, MA 02138, USA}

\author[0000-0002-8932-1219]{Ryan A. Loomis}
\affiliation{National Radio Astronomy Observatory, Charlottesville, VA 22903, USA}

\author[0000-0003-1413-1776]{Charles J. Law}
\affiliation{Center for Astrophysics \textbar\, Harvard \& Smithsonian, 60 Garden St., Cambridge, MA 02138, USA}

\author[0000-0001-6078-786X]{Catherine Walsh}
\affiliation{School of Physics and Astronomy, University of Leeds, Leeds, UK, LS2 9JT}

\author[0000-0003-4179-6394]{Edwin A. Bergin}
\affiliation{Department of Astronomy, University of Michigan,
323 West Hall, 1085 S. University Avenue,
Ann Arbor, MI 48109, USA}

\author[0000-0002-1637-7393]{Fran\c cois M\'enard}
\affiliation{Universit\'e Grenoble Alpes, CNRS, IPAG, F-38000 Grenoble, France}

\author[0000-0003-1526-7587]{David J. Wilner}
\affiliation{Center for Astrophysics \textbar\, Harvard \& Smithsonian, 60 Garden St., Cambridge, MA 02138, USA}

\author[0000-0003-2253-2270]{Sean M. Andrews}
\affiliation{Center for Astrophysics \textbar\, Harvard \& Smithsonian, 60 Garden St., Cambridge, MA 02138, USA}

\author[0000-0003-3283-6884]{Yuri Aikawa}
\affiliation{Department of Astronomy, Graduate School of Science, The University of Tokyo, 7-3-1 Hongo, Bunkyo-ku, Tokyo 113-0033, Japan}

\author[0000-0003-2014-2121]{Alice S. Booth}
\affiliation{School of Physics and Astronomy, University of Leeds, Leeds, UK, LS2 9JT}
\affiliation{Leiden Observatory, Leiden University, 2300 RA Leiden, the Netherlands}

\author[0000-0002-2700-9676]{Gianni Cataldi}
\affiliation{National Astronomical Observatory of Japan, 2-21-1 Osawa, Mitaka, Tokyo 181-8588, Japan}
\affiliation{Department of Astronomy, Graduate School of Science, The University of Tokyo, 7-3-1 Hongo, Bunkyo-ku, Tokyo 113-0033, Japan}

\author[0000-0002-8716-0482]{Jennifer B. Bergner} 
\altaffiliation{NASA Hubble Fellowship Program Sagan Fellow}
\affiliation{University of Chicago Department of the Geophysical Sciences, Chicago, IL 60637, USA}

\author[0000-0003-4001-3589]{Arthur D. Bosman}
\affiliation{Department of Astronomy, University of Michigan,
323 West Hall, 1085 S. University Avenue,
Ann Arbor, MI 48109, USA}

\author[0000-0003-2076-8001]{L. Ilse Cleeves}
\affiliation{University of Virginia, Charlottesville, VA 22903, USA}

\author[0000-0002-1483-8811]{Ian Czekala}
\altaffiliation{NASA Hubble Fellowship Program Sagan Fellow}
\affiliation{Department of Astronomy and Astrophysics, 525 Davey Laboratory, The Pennsylvania State University, University Park, PA 16802, USA}
\affiliation{Center for Exoplanets and Habitable Worlds, 525 Davey Laboratory, The Pennsylvania State University, University Park, PA 16802, USA}
\affiliation{Center for Astrostatistics, 525 Davey Laboratory, The Pennsylvania State University, University Park, PA 16802, USA}
\affiliation{Institute for Computational \& Data Sciences, The Pennsylvania State University, University Park, PA 16802, USA}
 \affiliation{Department of Astronomy, 501 Campbell Hall, University of California, Berkeley, CA 94720-3411, USA}

\author[0000-0002-2026-8157]{Kenji Furuya} \affiliation{National Astronomical Observatory of Japan, 2-21-1 Osawa, Mitaka, Tokyo 181-8588, Japan}

\author[0000-0003-4784-3040]{Viviana V. Guzm\'{a}n}
\affiliation{Instituto de Astrof\'isica, Pontificia Universidad Cat\'olica de Chile, Av. Vicu\~na Mackenna 4860, 7820436 Macul, Santiago, Chile}

\author[0000-0001-6947-6072]{Jane Huang}
\affiliation{Center for Astrophysics \textbar\, Harvard \& Smithsonian, 60 Garden St., Cambridge, MA 02138, USA}
\altaffiliation{NASA Hubble Fellowship Program Sagan Fellow}
\affiliation{Department of Astronomy, University of Michigan,
323 West Hall, 1085 S. University Avenue,
Ann Arbor, MI 48109, USA}

\author[0000-0003-1008-1142]{John~D.~Ilee} 
\affiliation{School of Physics and Astronomy, University of Leeds, Leeds, UK, LS2 9JT}

\author[0000-0002-7058-7682]{Hideko Nomura}
\affiliation{National Astronomical Observatory of Japan, 2-21-1 Osawa, Mitaka, Tokyo 181-8588, Japan}

\author[0000-0001-8642-1786]{Chunhua Qi}
\affiliation{Center for Astrophysics \textbar\, Harvard \& Smithsonian, 60 Garden St., Cambridge, MA 02138, USA}

\author[0000-0002-6429-9457]{Kamber R. Schwarz}
\altaffiliation{NASA Hubble Fellowship Program Sagan Fellow}
\affiliation{Lunar and Planetary Laboratory, University of Arizona, 1629 E. University Blvd, Tucson, AZ 85721, USA}

\author[0000-0002-6034-2892]{Takashi Tsukagoshi} \affiliation{National Astronomical Observatory of Japan, 2-21-1 Osawa, Mitaka, Tokyo 181-8588, Japan}

\author[0000-0003-4099-6941]{Yoshihide Yamato}
\affiliation{Department of Astronomy, Graduate School of Science, The University of Tokyo, 7-3-1 Hongo, Bunkyo-ku, Tokyo 113-0033, Japan}

\author[0000-0002-0661-7517]{Ke Zhang}
\altaffiliation{NASA Hubble Fellow}
\affiliation{Department of Astronomy, University of Wisconsin-Madison, 475 N Charter St, Madison, WI 53706, USA}
\affiliation{Department of Astronomy, University of Michigan, 323 West Hall, 1085 S. University Avenue, Ann Arbor, MI 48109, USA}

\begin{abstract}
Sulfur-bearing molecules play an important role in prebiotic chemistry and planet habitability. They are also proposed probes of chemical ages, elemental C/O ratio, and grain chemistry processing. Commonly detected in diverse astrophysical objects, including the Solar System, their distribution and chemistry remain, however, largely unknown in planet-forming disks. We present CS ($2-1$) observations at $\sim0\farcs3$ resolution performed within the ALMA-MAPS Large Program toward the five disks around IM Lup, GM Aur, AS 209, HD 163296, and MWC 480. CS is detected in all five disks, displaying a variety of radial intensity profiles and spatial distributions across the sample, including intriguing apparent azimuthal asymmetries. Transitions of \ce{C2S} and SO were also serendipitously covered but only upper limits are found. For MWC 480, we present complementary ALMA observations at $\sim0\farcs5$, of CS, $^{13}$CS, C$^{34}$S, \ce{H2CS}, OCS, and \ce{SO2}. We find a column density ratio N(\ce{H2CS})/N(CS)$\sim2/3$, suggesting that a substantial part of the sulfur reservoir in disks is in organic form (i.e., C$_x$H$_y$S$_z$). Using astrochemical disk modeling tuned to MWC 480, we demonstrate that $N$(CS)/$N$(SO) is a promising probe for the elemental C/O ratio. The comparison with the observations provides a super-solar C/O. We also find a depleted gas-phase S/H ratio, suggesting either that part of the sulfur reservoir is locked in solid phase or that it remains in an unidentified gas-phase reservoir. This paper is part of the MAPS special issue of the Astrophysical Journal Supplement.
\end{abstract}


\section{Introduction} \label{sec:intro}

Protoplanetary disks are a pivotal stage in the evolution from interstellar molecular clouds to planetary systems. Their chemical structures encode information both on the chemical evolution during star and planet formation, and on the future composition of planets. It is thus of fundamental importance to constrain and understand the chemistry of the principal chemical elements constituting these disks.
During the past decade, a myriad of studies focused on oxygen, carbon, and nitrogen chemistry in protoplanetary disks \citep[e.g.,][]{oberg2011_DISCS,guilloteau2016,kastner2018,cleeves2018,bergner2018,bergner2019,pontoppidan2019}, while
very little is known about sulfur chemistry in disks. This is probably because, more generally, the chemistry of sulfur in the Universe has remained a long-standing mystery for the past two decades \citep{ruffle1999,kama2019,navarro2020}.

Sulfur plays an important role in prebiotic chemistry \citep{chen2015} and planet habitability \citep{ranjan2018,ruf2019}. It is also one of the most abundant elements in the Universe with a solar value of S/H $\sim 1.5 \times 10^{-5}$~\citep{asplund2009}. In the diffuse interstellar medium (ISM) and photon-dissociation regions (PDR) the total amount of sulfur is close to the solar value \citep{goicoechea2006,howk2006}, while in dense molecular gas it is strongly depleted: less than $\sim 1\%$ of the sulfur solar abundance is observed in the gas phase \citep{tieftrunk1994,wakelam2004,vastel2018}. Therefore, a question yet to be answered is: what causes the observed sulfur depletion from diffuse to dense gas? While most of the sulfur is suspected to be locked into icy grain mantles \citep[e.g.,][]{millar1990,ruffle1999,vidal2017,laas2019}, only $\sim 4\%$ of the solar S-abundance has been detected in interstellar ices so far \citep{palumbo1997,boogert2015}. Therefore, the identity of the sulfur reservoir(s) in the ISM remains an open question.

In the Solar System, sulfur-bearing species are routinely detected, in the remnants of our own planet-forming disk such as comets, meteorites, and on planets and their satellites \citep[e.g.,][]{calmonte2016,hirschmann2016,lellouch2007,franz2019}. 
In particular, in comets, a dozen of S-bearing species have now been detected \citep{meier1997,bockelee-morvan2004,biver2016,calmonte2016}, including both complex S-molecules (\ce{CH3SH} and \ce{C2H6S}\footnote{Note that \ce{C2H6S} has two isomers, ethanethiol \citep[\ce{CH3CH2SH}, also known as ethyl mercaptan and only detected in Orion KL so far,][]{kolesnikova2014} and dimethyl sulphide (\ce{(CH3)2S}, that is, to our knowledge, not yet detected elsewhere in Space). However, the 67P/C-G measurements did not allow to distinguish the ratio of these two isomers.}) and multi-sulfuretted molecules, such as S$_2$, \ce{CS2}, \ce{S3}, and \ce{S4} which have not been detected yet, nor in the ISM, neither in protoplanetary disks. Studying the S-chemistry in disks is therefore crucial to understand the chemical origins of our own Solar System, and more generally, the role of sulfur in astrochemistry.

Among the approximately thirty different molecules detected in disks so far only five \footnote{For this inventory we did not include isotopologues, but note that the isotopologues $^{13}$CS and C$^{34}$S are also detected in disks~\citep{legal2019a,loomis2020}.} contain sulfur. 
These include CS, SO, \ce{H2S}, \ce{H2CS} and \ce{SO2} with the former two detected during the past two decades, and the latter three detected within the past few years due to the significant sensitivity improvements made in radio-interferometry instruments. CS is the most readily detected S-bearing species in proto-planetary disks  \citep{dutrey1997,fuente2010,guilloteau2016,teague2018,legal2019a}.
SO was the sole oxygenated, sulfur-bearing species detected in disks until the recent detection of \ce{SO2} in one disk \citep{booth2021}, which is probably indicative of a general highly reduced or O-poor gas chemistry in most disks. Another interesting point is that, so far, SO has only been detected toward a few young disks with signs of active accretion \citep{fuente2010,guilloteau2013,guilloteau2016,pacheco2016,booth2018,riviere2020}. \ce{H2S} has long been thought to be a main sulfur reservoir and is a major sulfur carrier in comet 67P/C-G \citep{calmonte2016}. However, it was only detected recently in the massive disk ($\sim$0.15~$M_\odot$) GG Tau with a \ce{H2S}/CS gas-phase column density ratio of $\sim 1/20$ \citep{phuong2018}, after unsuccessful searches in a handful of other disks \citep[GO Tau, MWC 480, DM Tau, and LkCa 15,][]{dutrey2011}. While additional \ce{H2S} observations in disks are required to draw firm conclusions, this result casts doubt on the importance of \ce{H2S} in disk gas-phase S-chemistry and has revived interest in the quest to identify the sulfur reservoir in disks \citep[e.g.,][]{kama2019}. The recent detection of \ce{H2CS} in the MWC~480 disk, with a \ce{H2CS}/CS gas-phase column density ratio of $\sim 1/3$, is in tension with recent models \citep{legal2019a} and suggests an incomplete theoretical understanding of disk S-chemistry. Thus, a better understanding of the S-chemistry is needed to inform astrochemical models and constrain the unseen reservoirs of S-bearing species, such as those locked onto icy dust grains, in disks.

Disks are vertically stratified into atmospheres, warm molecular layers, and cold midplanes, which are analogs to PDR, lukewarm molecular clouds, and cold dense cores, respectively \citep{aikawa2002,bergin2007,dutrey2014}. Recent sulfur-bearing species observations in each of these three types of astrophysical environments -- i.e., in a PDR \citep{fuente2017,riviere2019}, in a protostellar envelope \citep{drozdovskaya2018} and in dense cores \citep{vastel2018,navarro2020} -- have revived interest in the global quest for understanding the cycle of sulfur chemistry from molecular cloud to exoplanetary systems, and are timely for disk S-chemistry exploration. In this context, \cite{legal2019a} scrutinized the reactions pertinent to the sulfur chemistry within current gas-grain astrochemical models to constrain those molecules expected to be particularly abundant in disks, and predicted their radial and vertical distributions.

Here we present new observations of sulfur-bearing species in disks taken with the Atacama Large Millimeter/submillimeter Array (ALMA), as part of the Molecules with ALMA at Planet-forming Scales (MAPS) Large Program \citep{oberg2021}. The $^{12}$CS $J=2-1$ rotational transition was observed toward the five disks targeted within MAPS, i.e. the disks orbiting IM~Lup, GM~Aur, AS~209, HD~163296, and MWC~480 (for which stellar and disk properties are described in Table~\ref{tab:disk-properties}). The SO $J_{N}=2_3-1{_2}$ and $J_N=5_4-4_4$ and \ce{C2S} $J_N=8_7-7_6$ and $J_N=15_{14}-14_{15}$ rotational transitions were also serendipitously covered. In addition, we also present new complementary Cycle 6 ALMA observations (program 2018.1.01631.S, PI: R. Le~Gal) toward the MWC~480 disk, where $^{12}$CS $J=5-4$ and its $^{13}$CS and C$^{34}$S isotopologues were observed as well as several rotational transitions of \ce{H2CS}, OCS, and \ce{SO2}. 

The outline of the paper is as follows: we describe the observations in Section~\ref{sec:obs}, and we present the results, including the derivation of column densities and excitation temperatures, in Section~\ref{sec:obs-results}.
In Section~\ref{sec:modeling}, we present grids of disk chemistry models tuned to the MWC~480 disk where we obtained the most observational constraints. In Section~\ref{sec:discussion}, we discuss the observational and modeling results, and summarize our conclusions in Section~\ref{sec:conclusion}.

\section{Observations} \label{sec:obs}

\begin{deluxetable*}{lccccccccccc}
\tablecaption{Stellar and Disk Properties as presented in \cite{oberg2021} \label{tab:disk-properties}}
\tablehead{
\colhead{Source} & \colhead{Spectral Type} &
\colhead{dist.} & \colhead{incl} & \colhead{PA} & \colhead{T$_{\rm eff}$} &\colhead{$L_{\rm *}$}  &\colhead{Age\tablenotemark{a}}&\colhead{$M_*$\tablenotemark{b}}  &\colhead{log$_{10}$($\dot{M}$)} &\colhead{$v_{\rm sys}$}  &\colhead{References}\\
\colhead{} & \colhead{} 
&\colhead{[pc]}&\colhead{[$^{\circ}$]}&\colhead{[$^{\circ}$]}&\colhead{[K]} &\colhead{[$L_{\rm \odot}$]} & \colhead{[Myr]} &\colhead{[$M_{\rm \odot}$]} &\colhead{[$M_{\rm \odot}$ yr$^{-1}$]} & \colhead{[km~s$^{-1}$]}
}
\startdata
IM Lup  &K5 
&158    &47.5   &144.5  &4266   &2.57   & $0.2-1.3$ &1.1   &$-$7.9   & 4.5  & 1,2,3,4,5,6\\
GM Aur  &K6 
&159 &53.2  &57.2  &4350   &1.2     &$\sim2.5$& 1.1   &$-$8.1   & 5.6  & 1,7,8,9,10,11,12\\
AS 209  &K5 
&121    &35.0  &85.8  &4266   &1.41   &$\sim1$& 1.2    &$-$7.3 & 4.6 & 1,2,6,13,14\\
HD 162396   &A1   
& 101 &46.7   &133.3  &9332   &17.0 &$>5$ &2.0  &$-$7.4   &5.8 & 1,2,6,15,16\\
MWC 480 &A5 
&162   &37 &148    &8250   &21.9  &$\sim7$ &2.1   &$-$6.9   &5.1 & 1,17,18,19,20,21\\
\enddata
\tablenotetext{a}{The stellar ages are uncertain by at least a factor of two and should only be considered as preliminary estimates.}
\tablenotetext{b}{All stellar masses have been dynamically determined as described in \citet{teague2021}.}
\tablecomments{References are 1. \citet{gaia2018}; 2. \citet{huang2018b}; 3. \citet{alcala2017}; 4. \citet{pinte2018}; 5. \citet{mawet2012}; 6. \citet{andrews2018}; 
7. \citet{huang2020}; 8. \citet{macias2018}; 9. \citet{espaillat2010}; 10. \citet{kraus2009}; 
11. \citet{beck2019}; 12. \citet{ingleby2015}; 13. \citet{salyk2013}; 14. \citet{huang2017}; 15. \citet{fairlamb2015}; 16. \citet{teague2019a}; 17. \citet{liu2019}; 18. \citet{montesinos2009}; 19. \citet{simon2019}; 20. \citet{pietu2007}; 21. \citet{mendiguta2013}}
\end{deluxetable*}

\begin{table*}
\footnotesize
\begin{center}
\caption{List of Observations (molecular data from CDMS$^{(a)}$) \label{tab:obs-list}}
 \renewcommand{\arraystretch}{1.2}
\begin{tabular}{lcccclccccc}
\hline\hline
Species&Transition&Frequency&$E_{u}$&Log$_{10}(A_{\rm ij})$ &Source& RMS$_{\rm{chan}}$& \multicolumn{2}{c}{Restored Beam}&$R_{\rm{max}}^{(b)}$&$S_\nu\Delta_v$($R_{\rm{max}}^{(c)}$)\\
&&(GHz)&(K)&(s$^{-1}$)&&(mJy/beam)&($''\, \times \,''$)&(\degr)&($''$)&(mJy km/s)\\ 
\hline
\multicolumn{11}{c}{MAPS data (Project ID: 2018.1.01055.L)}\\
\hline
$^{12}$CS& $2-1$&97.98095 &7.1 & $-4.7763$ &  IM~Lup&0.51&$0.30\times0.23$&$-80.2$&  $3.0\pm0.1$ &$268\pm11$\\ 
&&&&& GM~Aur&0.46&$0.39\times0.27$&5.1& $2.1\pm0.1$ &$133\pm 3$\\
&&&&&AS~209&0.48&$0.33\times0.26$&$-66.7$&$0.9\pm0.1$ &$166\pm4$\\ 
&&&&& HD~163296&0.42&$0.31\times0.24$&$-88.2$& $1.5\pm0.1$&$302\pm 15$\\
&&&&& MWC~480&0.46&$0.39\times0.28$&7.2& $2.0\pm0.1$&$48\pm 4$\\
\hline
SO& $2_3-1_2$&99.29987 &9.2 & $-4.9488$ &  IM~Lup&0.44&$1.23\times1.12$&79.3& $3.0\pm0.1$ &$\lesssim 30 $\\
&&&&& GM~Aur&0.42&$1.44\times1.29$&$-5.9$& $2.1\pm0.1$ &$< 10$\\
&&&&&AS~209&0.40&$1.33\times1.11$&82.7&$0.9\pm0.1$&$< 15$\\
&&&&& HD~163296&0.39&$1.22\times1.05$&85.3&$1.5\pm0.1$&$< 45$\\
&&&&& MWC~480&0.43&$1.44\times1.29$&$-3.6$& $1.5\pm0.2$& $\lesssim14$\\
& $5_4-4_4$&100.02964 &38.6 & $-5.9656$ &  IM~Lup&0.50&$1.22\times1.09$&82.1& $3.0\pm0.1$ &$< 51 $\\
&&&&& GM~Aur&0.45&$2.10\times1.37$&$-29.0$& $2.1\pm0.1$ &$\lesssim4$\\
&&&&& AS~209&0.41&$1.28\times1.10$&$-80.9$&$0.9\pm0.1$ &$<18$\\
&&&&& HD~163296&0.41&$1.26\times1.07$& 85.0 & $1.5\pm0.1$ &$<53 $\\
&&&&& MWC~480&0.41&$2.08\times1.38$&$-28.8$& $1.5\pm0.2$ &$<6$\\
\hline
\ce{C2S}& $8_7-7_6$& 99.86652& 28.1 & $-4.3562$ &  IM~Lup&0.49&$1.22\times1.09$& 82.1 & $3.0\pm0.1$ &$< 51 $\\
&&&&& GM~Aur&0.45&$2.10\times1.37$&$-29.0$& $2.1\pm0.1$ &$<9$\\
&&&&& AS~209& 0.41&$1.29\times1.10$& $-80.9$&$0.9\pm0.1$ &$\lesssim 9$\\
&&&&& HD~163296&0.39&$1.26\times1.07$&85.0& $1.5\pm0.1$ &$< 53$\\
&&&&& MWC~480&0.48&$2.08\times1.38$&$-28.8$& $1.5\pm0.2$ &$< 5$\\
\hline
\multicolumn{11}{c}{Complementary Cycle 6 unpublished ALMA data (project ID: 2018.1.01631.S)}\\
\hline
$^{12}$CS& $5-4$ &244.93564 &35.3& $-3.5271$ &MWC~480&3.9&$0.71\times0.45$&$-12.9$&$1.5\pm0.2$&$98 \pm 5$\\
\hline
$^{13}$CS& $5-4$ & 231.220996 & 33.3 & $-3.6008$ & MWC~480 &4.4 & $0.83\times0.58$ & 0.7& $1.5\pm0.2$ &$<7$\\
\hline
C$^{34}$S& $5-4$ &241.016088 &27.8 &$-3.5568$& MWC~480&3.4&$0.86\times0.59$&6.8& $1.5\pm0.2$ &$20\pm5$\\
\hline
H$_2$CS& $7_{16}-6_{15}$&244.0485044& 60.0 & $-3.6771$&	 MWC~480&3.2&$0.79\times0.55$& 1.0 &$1.5\pm0.2$ &$ 29\pm 5$\\
& $7_{26}-6_{25}$ & 240.3820512  &   98.8 &$-3.7248$ & & 3.9 &$0.87\times0.59$&6.8& &$ <13$\\
& $7_{35}-6_{34}$ & 240.3930370 & 164.6 & $-3.7760$ &  &4.0&$0.86\times0.59$&6.3& &$ <10$\\
& $7_{34}-6_{33}$ & 240.3937618  & 164.6 & $-3.7760$ & &4.0&$0.86\times0.59$&6.3& &$ <10$\\
\hline
OCS& $19-18$ & 231.06099 & 110.9 &$-4.4463$&  MWC~480&3.3&$0.78\times0.54$&$-0.02$&$1.5\pm0.2$ & $ < 5$\\
& $20-19$ & 243.21804 &  122.6 & $-4.3790$ & &3.0&$0.79\times0.55$&0.76& & $< 13$\\
\hline
\ce{SO2}& $11_{57}-12_{48}$ & 229.3476299 & 122 & $-4.7194$ & MWC~480& 2.7 &$0.84\times0.58$& 1.4 & $1.5\pm0.2$ & $18\pm6$\\
&   $5_{24}-4_{13}$ & 241.6157967 & 23.6 & $-4.0728$ & &3.2&$0.86\times0.59$&5.9 & &$ < 6$\\
& $5_{42}-6_{33}$ & 243.0876473 & 53.1 & $-4.9886$ & &2.9&$0.79\times0.55$&1.2 & &$ < 6$\\
\hline
\end{tabular}
\end{center}
\tablenotetext{a}{\texttt{https://cdms.astro.uni-koeln.de/cdms/portal/}, \cite{cdms2001,cdms2005}}
\tablenotetext{b}{$R_{\rm{max}}$ stands for the outer radius of the molecular line emission where 90\% of the cumulative flux from the radial profiles is contained. The uncertainty is $1\sigma$ error.}
\tablenotetext{c}{$S_\nu\Delta_v$($R_{\rm{max}}$) corresponds to the flux density integrated out to the outer radius $R_{\rm{max}}$ of the molecular line emission.}
\end{table*}

We used three sets of ALMA observational data. First, new observations obtained with MAPS (program number: 2018.1.01055.L, PI: K. \"Oberg) of the $^{12}$CS $2-1$ rotational transition, and of two rotational transitions of SO and \ce{C2S} that were serendipitously covered in the same program. Second, new observations obtained with another ALMA program (program number: 2018.1.01631.S, PI: R. Le~Gal) of CS, $^{13}$CS, and C$^{34}$S $5-4$, four \ce{H2CS} rotational transitions, two OCS rotational transitions, and three \ce{SO2} rotational transitions. Finally, to get better estimates of the column densities and exitation temperatures of \ce{H2CS}, $^{12}$CS, and C$^{34}$S, we also used already published complementary ALMA data of additional detected rotational transitions of these molecules \citep{legal2019a}. The new observations are further described below and their molecular transitions, their frequencies, and spectroscopic parameters are listed in Table~\ref{tab:obs-list}.

\subsection{MAPS observations}
\label{subsec:description_obs_MAPS}

The CS $2-1$ rotational transition was observed in the five MAPS disks in Band 3 with an angular resolution of $\sim 0\farcs3$ (see Table~\ref{tab:obs-list}) and a spectral resolution of 71~kHz, corresponding to $\sim 0.22$~km/s at 97~GHz. 
More details about the observations can be found in \cite{oberg2021}. For the descriptions of the reduction and imaging procedures applied to the CS $2-1$ MAPS observations, we refer the reader to \cite{oberg2021} and \cite{czekala2021}, respectively. Here we used the CS $2-1$ images created with a robustness parameter of 0.5 for the Briggs weighting which results in slightly higher resolution images than the fiducial images presented in \cite{oberg2021} and \cite{law21_rad} which used circularized $0\farcs3$ beams.

While SO and \ce{C2S} were not part of the main targeted molecules within the MAPS program, two of their rotational transitions -- namely the $2_3-1_2$ (at 99.29987~GHz) and $5_4-4_4$ (at 100.2964~GHz) transitions for SO, and the $8_7-7_6$ (at 99.86652~GHz) and $15_{14}-14_{15}$ (at 234.81596~GHz) lines for \ce{C2S} -- were covered in Band 3 and 6, at lower spectral resolution (1.129~MHz, i.e. $\sim 3.4$~km/s at 100~GHz and $\sim 1.4$ km/s at 235~GHz).
After continuum subtraction with the CASA \texttt{uvcontsub} function, we \texttt{CLEAN}ed \citep{hogbom1974} the \ce{C2S} and SO data using the same procedure as outlined in \cite{czekala2021}. As these lines are expected to be weak, we applied a robustness parameter of 1 and $1''$ {\it uv}-taper to improve their imaging and signal-to-noise ratio (SNR). All MAPS images used here are available for download through the ALMA Archive via \url{https://almascience.nrao.edu/alma-data/lp/maps}. An interactive browser for this repository is also available on the MAPS project homepage at \url{http://www.alma-maps.info}.

\subsection{Complementary ALMA observations of MWC~480}\label{subsec:description_obs_C6data}

Independently from the MAPS program, Cycle 6 ALMA observations toward the MWC 480 disk were conducted on 2019 April 30 in three execution blocks (EB) with an angular resolution of $\sim0\farcs55$, as part of program 2018.1.01631.S (PI: R. Le~Gal). We are presenting and using these data here to complement the data-set of sulfur-bearing molecules observed for this disk. The measurements used ALMA Band 6 receivers, with correlated data 
divided into thirteen spectral windows (SPWs). SPWs were centered on twelve different rotational transitions of sulfur-bearing molecules, including the CS, $^{13}$CS, and C$^{34}$S $5-4$ lines, four \ce{H2CS} lines, two OCS lines, and three \ce{SO2} lines. Each SPW contains 480 channels with a total bandwidth of 58.59~MHz, with a 0.141~MHz resolution per channel, corresponding to a velocity resolution of $\sim$~0.18~km/s. 
One SPW was reserved for high sensitivity continuum observations to aid in the self-calibration of the data. The total on-source integration time was $\sim 43$ minutes. A total of 42 and 43 antennas were included for the first EB and the remaining two EBs, respectively, and covered baselines from 15 to 740~m. All EBs used the source J0510+1800 as their
bandpass and flux calibrators and the source J0438+3004 as phase calibrator. Only one third of the proposed observations were performed, and both the RMS and beam size failed to meet our requested performance parameters. Therefore, the observations were deemed to QA2 SEMI-PASS and the data released by the observatory. However, the data quality already allows us to derive constraints on the S-chemistry as described below.

Data calibration was initially pipeline-processed. We then use the Common Astronomy Software Application package (CASA) version CASA 5.6.1-8 \citep{mcmullin2007} to reduce the data. 
Self-calibration was performed using the SPW reserved for continuum. We performed three iterations of phase self-calibrations, and then one amplitude self-calibration.
After continuum subtraction with the CASA \texttt{uvcontsub} function, the data were \texttt{CLEAN}ed \citep{hogbom1974} using $3\sigma$ noise threshold and Briggs weighting with a robustness parameter of 0.5 for the main CS isotopologue and of 1 with a taper of 1$''$ for the other lines to improve their imaging and SNR.
The RMS per channel of all the observations presented in this study are listed in Table~\ref{tab:obs-list}.

\section{Observational Results} \label{sec:obs-results}

\begin{figure*}
\centering
\includegraphics[scale=0.41]{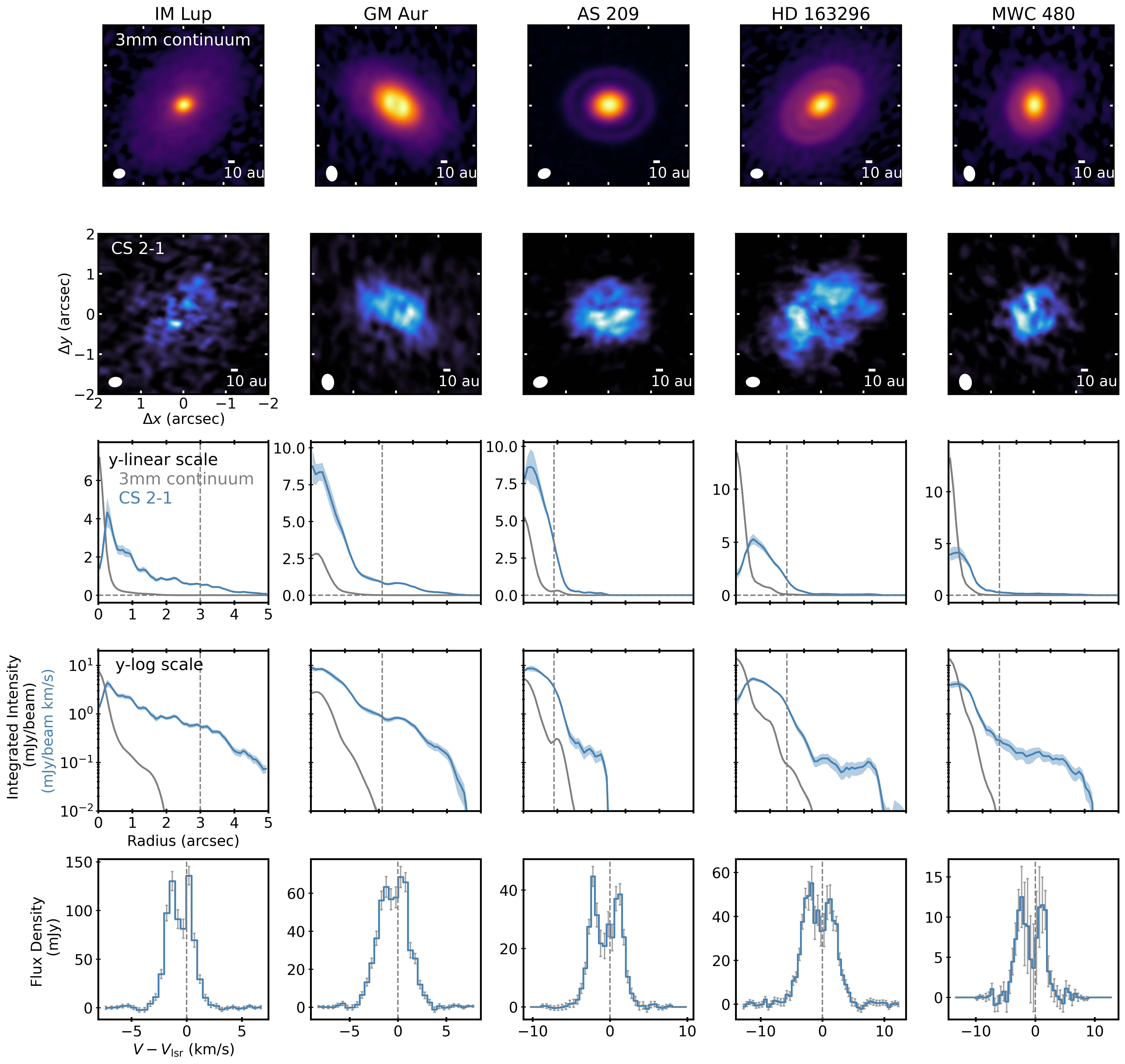}
\caption{Zeroth moment maps and radial intensity profiles for the MAPS disk sample, ordered by increasing stellar mass, see Table~\ref{tab:disk-properties}, from left to right. {\it First row:} Zeroth moment maps of the dust continuum at 3~mm produced using an arcsinh color stretch for the AS~209 disk, and a power-law color stretch for all four other disks, to enhance the faint and extended emission.
{\it Second row:} Zeroth moment maps of
    CS $2-1$.
    Synthesized beams are shown in the
    lower left corner of each panel. 
    {\it Third and fourth rows:} Radially de-projected and azimuthally averaged intensity profiles of the continuum and the CS $2-1$ emission in y-linear and y-log scales, respectively. The vertical dashed gray lines indicate the  outer  radius $R_\mathrm{max}$ of the molecular line emission where 90\% of the cumulative flux from the radial profiles is contained within 1$\sigma$ error.
    {\it Fifth row:} Integrated intensity spectra of CS $2-1$. The uncertainties on radial and intensity profiles are calculated as the standard deviation on the mean in the radial annulus over which the emission was averaged, following the MAPS collaboration convention, described in detail in \cite{law21_rad}. So, these error bars do not include the absolute calibration uncertainty of 10\%}.
    \label{fig:MAPS-CS21-obs-overview}
\end{figure*}

\subsection{CS 2-1 fluxes and spatial distributions}
\label{subsec:CS-morpho}

Figure~\ref{fig:MAPS-CS21-obs-overview} displays the integrated intensity (zeroth moment) maps of the spatially resolved MAPS observations of the CS $2-1$ rotational transition toward each of the five targeted disks. To build these maps we used the Python package \texttt{bettermoments} \citep{teague2018_bettermom} applied to the image cube available for download  in  the  MAPS  data  repository. We used an hybrid mask combining a Keplerian mask (also available for download  in  the  MAPS  data  repository) and $3 \sigma$ clip to mask any pixels below this threshold. For comparison, we also show the 3~mm continuum emission maps made in \cite{sierra2021}. 
Radially de-projected and azimuthally averaged intensity profiles of the continuum and CS $2-1$ line are also shown for each of the five MAPS disks. These were produced using the \texttt{radial profile} function of the Python package \texttt{GoFish} \citep{teague2019}, considering the disk physical parameters (i.e., disk inclination, disk position angle, mass of the central star, and distance) listed in Table~\ref{tab:disk-properties}. The angular resolution is $\sim 0.3''$, i.e., ranging from 30 au (HD~163296) to 49~au (MWC~480) depending on distance across the sample of MAPS disks.
Finally, the CS 2-1 spectra are also depicted in Fig.~\ref{fig:MAPS-CS21-obs-overview} for each targeted disks, showing a typical double-peaked profile indicative of the Keplerian rotation of the disk. 

Based on the radial intensity profiles of the CS $2-1$ emission across the disk sample, central holes appear for IM Lup and HD 163296, with the largest radial hole extent found toward HD~163296. For the other three disks, the SNR and spatial resolution are not sufficient to infer the morphology of the inner disk emission.
Beyond the inner holes we see a wide diversity in the morphology and extent of the CS $2-1$ radial intensity profiles compared to the dust continuum. For instance, emission plateaus appear for IM~Lup, GM~Aur, and HD~163296, leading to outer CS $2-1$ emission radii a factor of $\approx$~2 larger than the dust continuum. 
The GM~Aur disk -- the only transition disk in the MAPS sample -- is the only disk in which we see a tentative outer emission ring at $\sim 2\farcs5$ (i.e. $\sim400$~au). 

\begin{figure*}
    \includegraphics[scale=0.49]{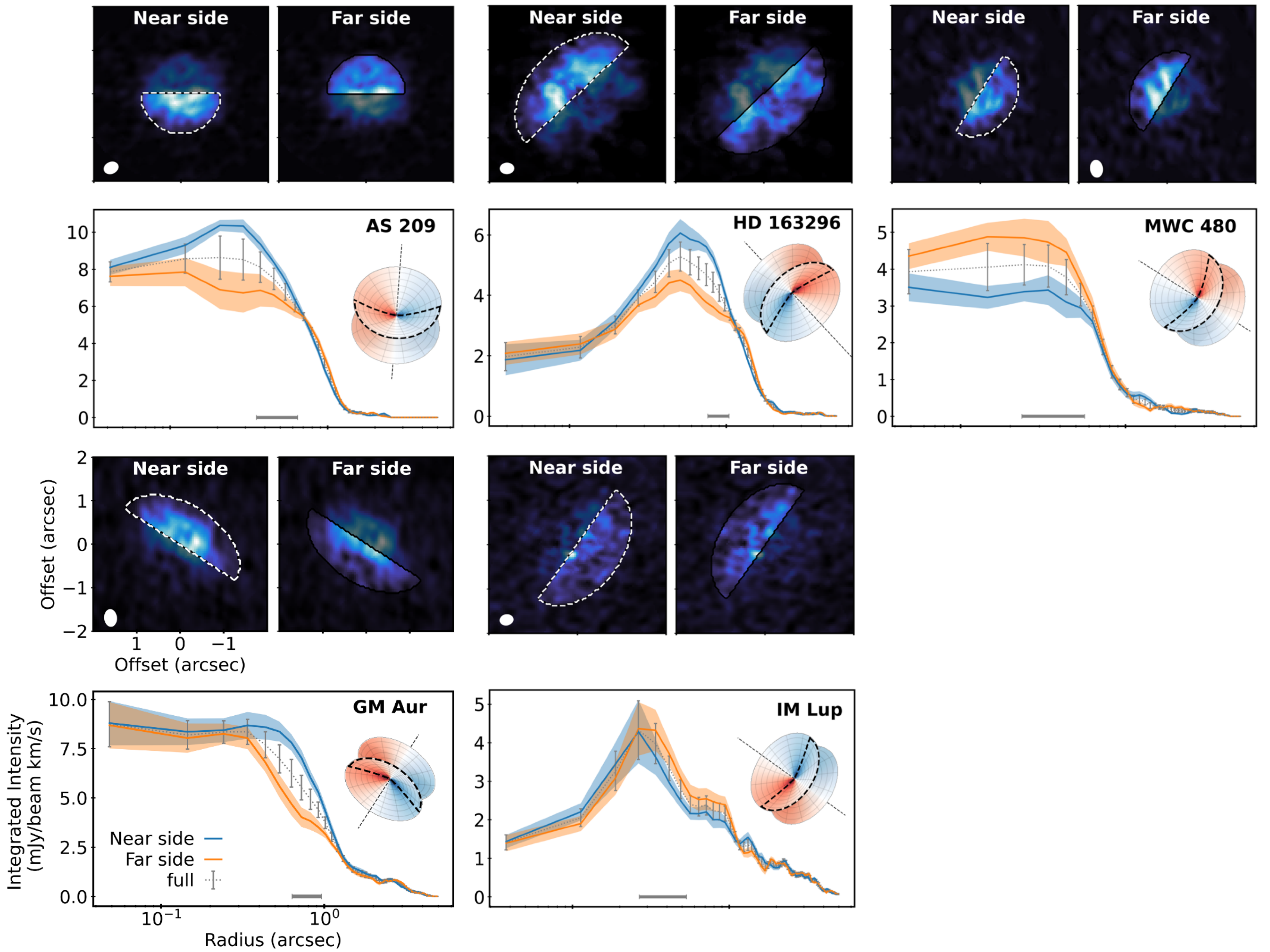}
    \caption{{\it Upper and third rows:} Zeroth moment maps of the CS 2-1 emission in the far and near sides of each of the five MAPS disks. Synthesized beams are shown in the lower left corner of the panels presenting the far side of each disk. {\it Second and lower row:} Radially  de-projected  and averaged intensity profiles of the furthest (blue) and near (orange) CS 2-1 emission sides of each disks compared to the azimuthally averaged intensity profiles (gray dotted line). The uncertainties on the radial intensity profiles are calculated as the standard deviation on the mean in the radial annulus over which the emission was averaged. Synthesized beams are shown by the gray error bar below each radial intensity profiles. The insets represent a schematic view of each disk inclination.}
    \label{fig:asym-disks}
\end{figure*}

Interestingly, the zeroth moment maps of the CS $2-1$ emission show some asymmetries that do not appear in the dust emission. In particular, toward four of the sources (GM~Aur, AS~209, HD~163296, and MWC~480), we find up to a factor 2 or $5 \sigma$ of intensity difference between the near and far sides of the disks. Since the five MAPS disks have a non-zero inclination (see Table~\ref{tab:disk-properties}), the closest and farthest half disk sides, with respect to the disk semi-major axis, are defined as near and far disk sides, respectively. This is illustrated by the schematic views of the geometry of each disk that we show as insets in Fig.~\ref{fig:asym-disks}. For three of the five MAPS sources -- namely AS~209, HD~163296, and GM~Aur -- the brightest CS emission sides coincide with the near side of the disks. Intriguingly, the reverse is observed for MWC~480, where the brightest CS emission side coincides with the far side of the disk. However, given the relatively low SNR, the robustness of these asymmetries is hard to assess. As for IM~Lup, which is the disk with the smallest CS 2-1 integrated intensity, we do not observe such asymmetries. These asymmetries are further discussed in Sect.~\ref{subsec:CS-asym}.

\begin{figure}
    \centering
        \includegraphics[scale=0.36]{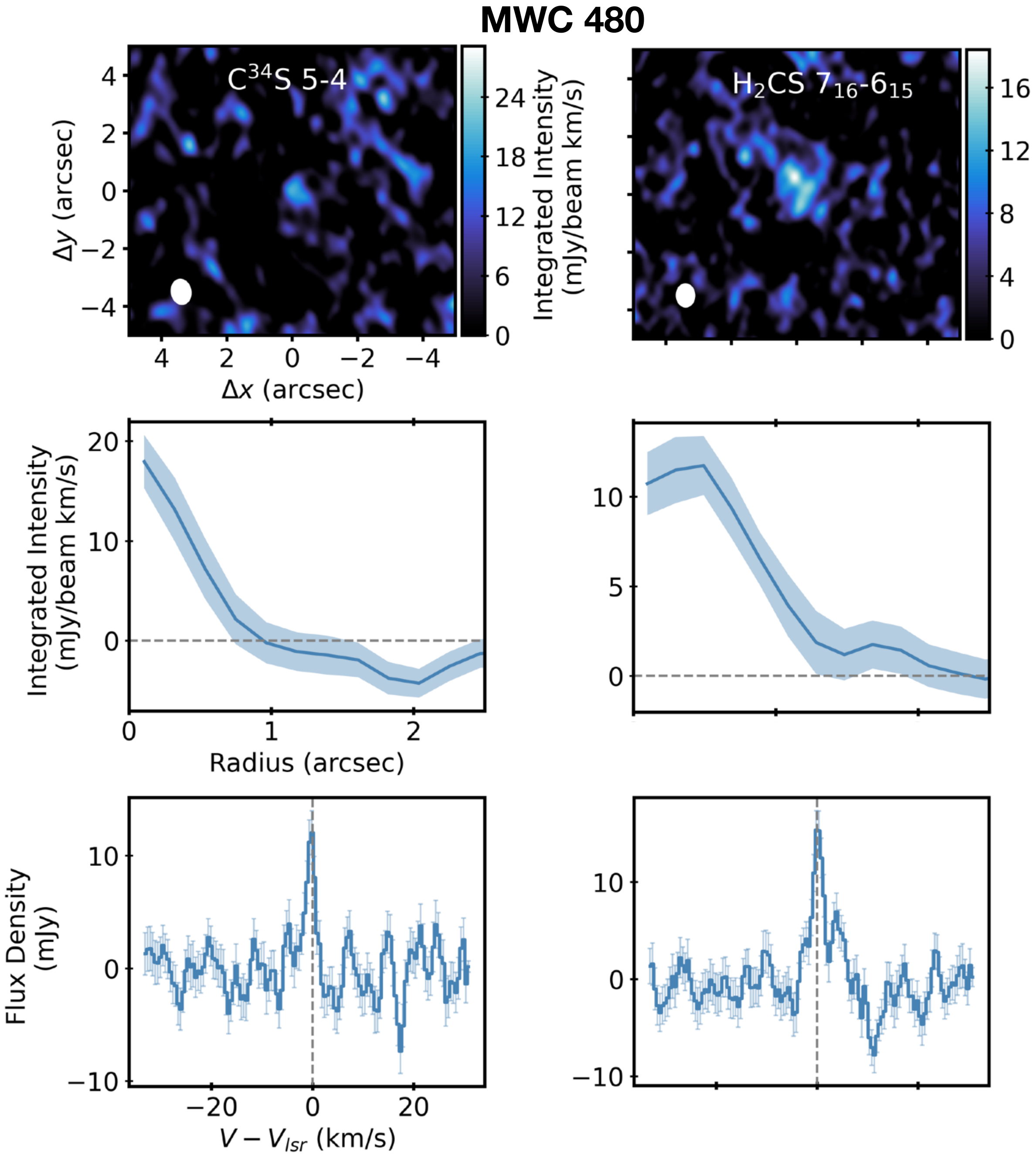}
    \caption{Zeroth moment maps (top panels), radially de-projected and azimuthally averaged intensity profiles within $1\sigma$ - as in Fig.~\ref{fig:MAPS-CS21-obs-overview} - (middle panels), and shifted and stacked disk-integrated line spectra of the C$^{34}$S $5-4$ (left) and \ce{H2CS} $7_{16}-6_{15}$ (right) rotational transitions observed toward the MWC~480 disk with $1\sigma$ (bottom panels). These last  uncertainties are calculated on a per channel basis, taking into account de-correlation along the spectral axis \citep[see also][]{yen2016,ilee2021}.}
    \label{fig:C34S-H2CS-mwc480-mom-rad-spec}
\end{figure}

\subsection{CS isotopologues and \ce{H2CS} in MWC 480}
\label{subsec:obs-CS-isotopologues-and_H2CS}

The $^{13}$CS and C$^{34}$S $5-4$ rotational transitions were observed as part of our complementary ALMA program toward MWC~480 (see Sect.~\ref{subsec:description_obs_C6data}). We did not detect the $^{13}$CS $5-4$ line neither with matched filtering method \citep[\texttt{VISIBLE},][]{loomis2018_match_filter} nor with line velocity shift and stacking techniques~\citep[\texttt{GoFish},][]{teague2019}. The latter exploits the known geometry and velocity structure of the disk to de-project the rotation profile and combine Doppler shifted emission to a common centroid velocity reference frame. This results in a single disk-integrated spectrum for each transition. 
However, imaging the C$^{34}$S $5-4$ line reveals a $\sim 3-4\sigma$ detection that is shown in Fig.~\ref{fig:C34S-H2CS-mwc480-mom-rad-spec} and reported in Table~\ref{tab:obs-list}. This detection is confirmed when we build the integrated spectrum of the line using the velocity shift and stacking methods of \texttt{GoFish} (see bottom panel in~Fig.~\ref{fig:C34S-H2CS-mwc480-mom-rad-spec}). 
These results are consistent with the $^{13}$CS and C$^{34}$S $6-5$ observations reported in \cite{legal2019a}, where the C$^{34}$S line was tentatively detected toward the MWC~480 disk while $^{13}$CS was not.

Four \ce{H2CS} transitions (see Table~\ref{tab:obs-list}) were also observed as part of our complementary ALMA program toward the MWC~480 disk. Among these four transitions, only the \ce{H2CS} line with the lowest upper energy level (i.e., \ce{H2CS} $7_{16}-6_{15}$) is detected, with a $\sim 5-6\sigma$ detection. The zeroth moment map, the radially de-projected and azimuthally averaged intensity profile, and the shifted and stacked disk-integrated spectrum of the \ce{H2CS} $7_{16}-6_{15}$ detection are shown in Fig.~\ref{fig:C34S-H2CS-mwc480-mom-rad-spec}. 
As for the remaining three \ce{H2CS} lines, their non-detections are not surprising regarding their upper energy levels and line strengths (see Table~\ref{tab:obs-list}, where we also report upper limits in Table~\ref{tab:obs-list}).

\subsection{Multi-line analysis}
\label{subsec:multi_CS_analysis}

To constrain the $^{12}$CS, C$^{34}$S, and \ce{H2CS} column densities and excitation temperatures toward the MWC~480 disk, we combined the new observations presented here with complementary ALMA observations of $^{12}$CS ($5-4$ and $6-5$), C$^{34}$S ($6-5$), and \ce{H2CS} ($8_{17}-7_{16}$, $9_{19}-8_{18}$, and $9_{18}–8_{17}$), already published in \cite{legal2019a}. Assuming optically thin lines and local thermal equilibrium (LTE), we used a rotational diagram analysis \citep{goldsmith1999} to derive the disk-integrated column densities and excitation temperatures of these molecules. These quantities are derived from the disk-integrated flux densities, as described in \cite{legal2019a} and summarized below.

\begin{figure*}
    \centering
        \includegraphics[scale=0.65]{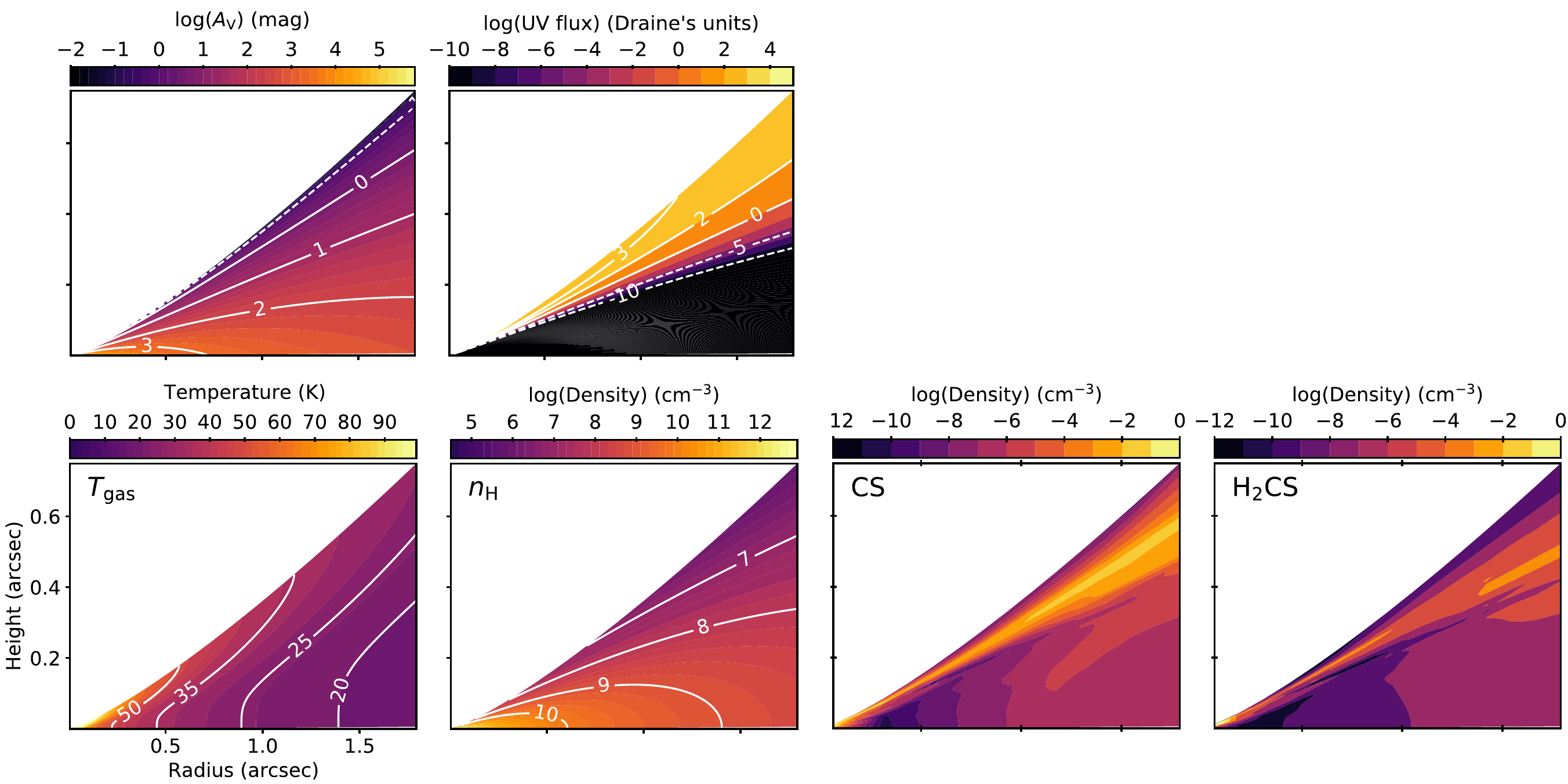}
    \caption{\underline{First row:} Visual extinction and UV flux profiles fed in our MWC~480 protoplanetary disk astrochemical model \citep{legal2019a}.
    \underline{Second row:} First two panels show the temperature and density 2D profiles fed in our MWC~480 protoplanetary disk astrochemical model. The third and fourth panels show the modeled number densities (i.e., absolute abundances) of CS, \ce{H2CS}. All panels are represented as functions of disk radius vs. height.}
    \label{fig:T-n-profiles_and_CS_H2CS}
\end{figure*}

To illustrate that the LTE assumption is justified we show in Figure~\ref{fig:T-n-profiles_and_CS_H2CS} the profiles of the main physical parameters used to build the MWC~480 disk physical structure along with the number density computed for CS and \ce{H2CS} with our corresponding published model \cite{legal2019a}. This allows us to show that the gas density in the main emitting molecular layers is larger than $10^7\ccc$, i.e. well above the critical densities of the observed CS and \ce{H2CS} transitions, which justifies well the LTE assumption. For temperatures in the range $20-50$~K, the critical densities for CS are in the ranges $\sim 7\times10^4 - 3\times10^6 \ccc$ \citep{shirley2015}, and in the range $2 - 4 \times10^6\ccc$ for \ce{H2CS}, using scaled \ce{H2CO} collisional rates from \cite{wiesenfeld2013} \citep[see the \texttt{Leiden Atomic and Molecular Database (LAMDA)\footnote{https://home.strw.leidenuniv.nl/~moldata/},}][]{vandertak2020}.

Assuming optically thin transitions, the disk-integrated flux densities $S_\nu \Delta v$, can be related to the column density of their respective upper energy state, $N_u$, as follows:
\begin{equation}
    N_u= \frac{4\pi S_\nu \Delta v}{A_{ul} \Omega hc},
\label{eq:Nu}
\end{equation}
where $S_\nu$ is the flux density, $\Delta v$ the line width, $A_{ul}$ the Einstein coefficient, $c$ the speed of light, and
$\Omega$ the solid angle subtended by the source \citep[e.g.,][]{bisschop2008,loomis2018_ch3cn}. For this analysis, we use the disk flux densities $S_\nu \Delta v$ integrated out to the outer radius of the molecular line emission, referred to as R$_{\rm{max}}$ in Table~\ref{tab:obs-list}.

The disk-integrated column density, $N_{\rm{tot}}$, and excitation temperature, $T$, can then be derived from the upper level population, $N_u$, which follows the Boltzmann distribution:
\begin{equation}
   N_{\rm{tot}}= \frac{N_u}{g_u} Q_{{\rm{rot}}}(T)e^{E_u/k_BT},
    \label{eq:boltzmann_dist}
\end{equation}
with $g_u$ and $E_u$ the degeneracy and energy of the upper energy level $u$, respectively; $k_{\rm{B}}$ the Boltzmann constant; and $Q_{\rm{rot}}$ the partition function of the molecule, which for a diatomic molecule such as CS can be approximated by:
\begin{equation}
    Q_{\rm{rot}}(T) \approx \frac{k_B\,T}{h\,B_0}+\frac{1}{3}.
    \label{eq:part_func}
\end{equation}
In this expression $h$ is the Planck constant and $B_0$ is the rotational constant of the molecule. For CS we used $B_0 = 24495.562\times10^6$~Hz (see CDMS).
For \ce{H2CS} we interpolated the \{$T,Q_{\rm{rot}}(T)$\} values provided by CDMS.

Using Eq.~\ref{eq:boltzmann_dist} and appendix B of \cite{legal2019a}, the optical depth of a given transition at temperature $T$ can be expressed as:

\begin{equation}
   \tau_\nu = \sqrt{\frac{4\ln2}{\pi}} \frac{ N_u A_{ul} \, c^3}{\Delta v_{\rm{FWHM}} \, 8\pi \nu^3}(e^{h\nu/k_B\tex} -1),
   \label{eq:int_tau_nu_gauss_prof}
\end{equation}
where $\Delta v_{\rm{FWHM}}=\sqrt{8\ln 2}\,\sigma_v$ is the full width at half maximum of the observed transition. $\sigma_v$ is the width of the Gaussian fit, since for optically thin lines, the line profiles remain Gaussian. 

As described in appendix B of \cite{legal2019a}, we can substitute Eq.~\ref{eq:int_tau_nu_gauss_prof} in $C_\tau =\frac{\tau}{1-e^{-\tau}}$, which corresponds to the "optical depth correction factor" for a square line profile in case $\tau~\cancel{\ll}~1$ \citep{goldsmith1999}. This allows us to 
build a likelihood function $\mathcal{L}({\rm data},N_{\rm{tot}},\tex)$ that we used with the Python implementation \texttt{emcee} \citep{emcee2013} of the affine-invariant ensemble sampler for Markov Chain Monte Carlo (MCMC) \citep{goodman2010} to compute posterior probability distributions for \tex\ and $N_{\rm {tot}}$. The following uniform and permissive priors were assumed:
\begin{eqnarray}
    \tex (\rm{K}) = \mathcal{U}(3,300)\\
    N_{\rm {tot}} (\cc) = \mathcal{U}(\dix{7},\dix{20}).
\end{eqnarray}

\subsection{Disk-integrated column densities in MWC~480}
\label{subsec:disk-int-col-dens-MWC480}
\begin{figure}
    \centering
    \includegraphics[scale=0.7]{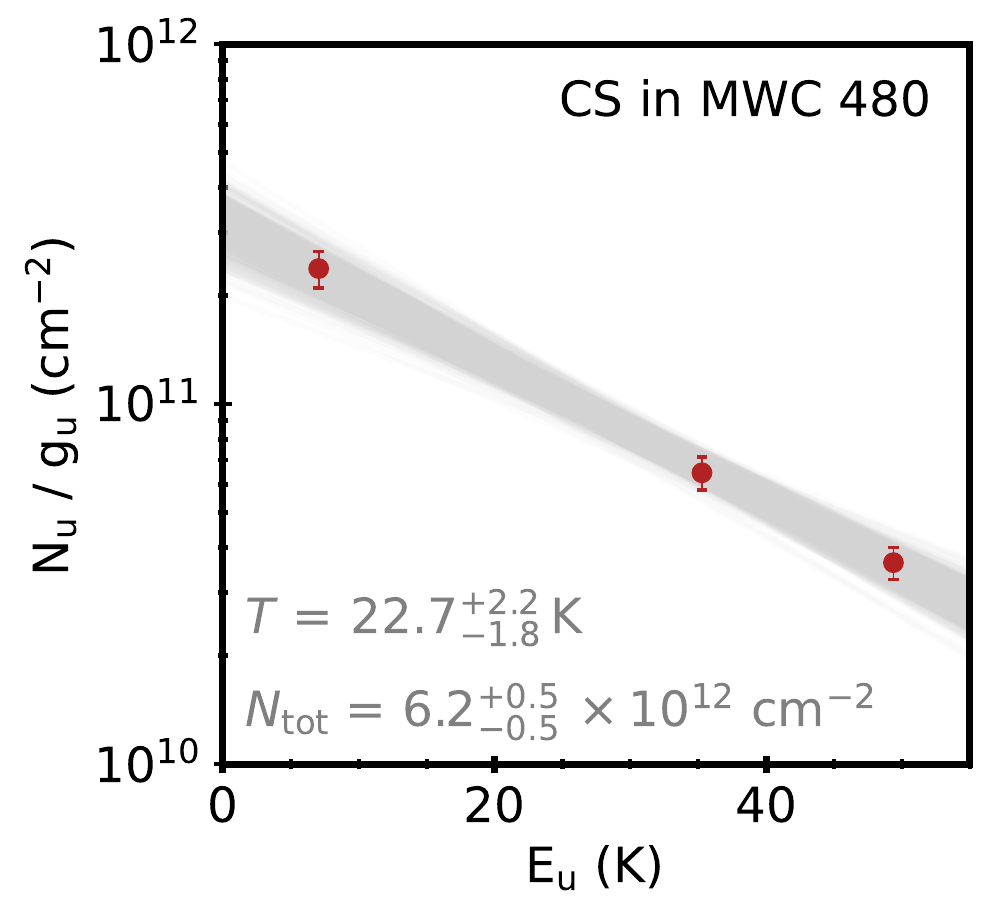}\\
    \includegraphics[scale=0.7]{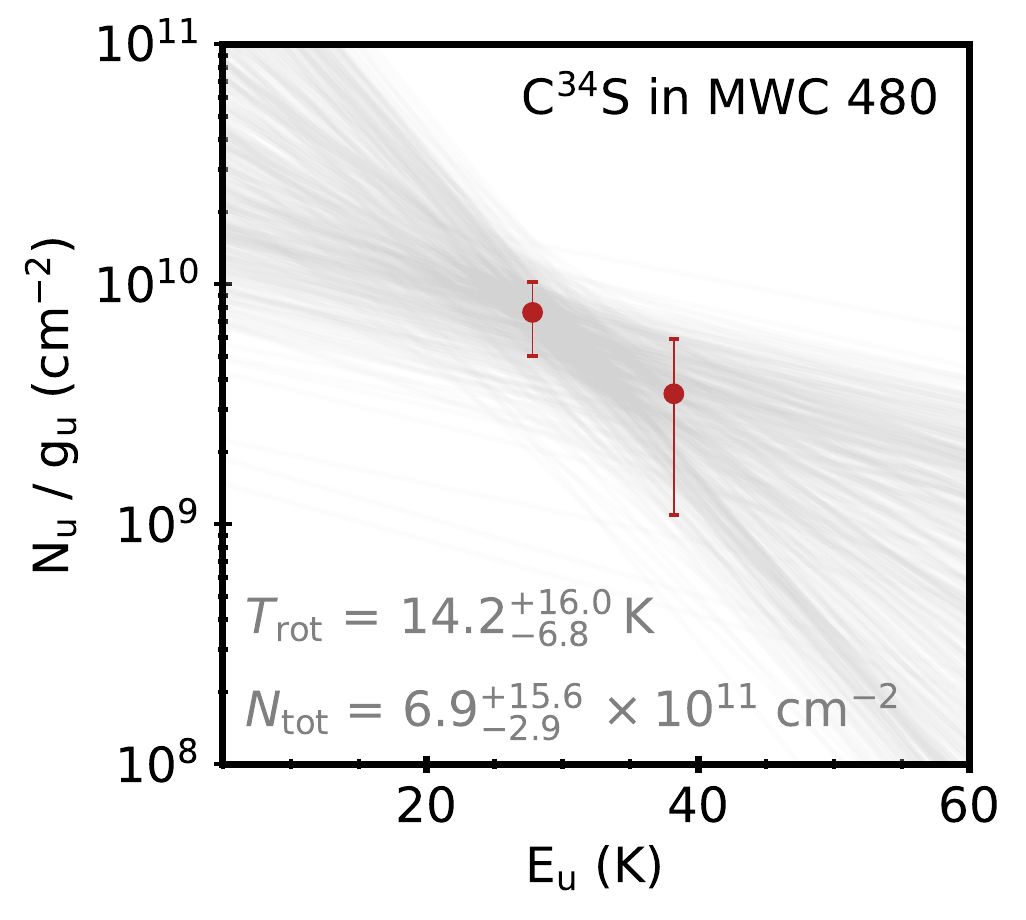}\\
    \includegraphics[scale=0.7]{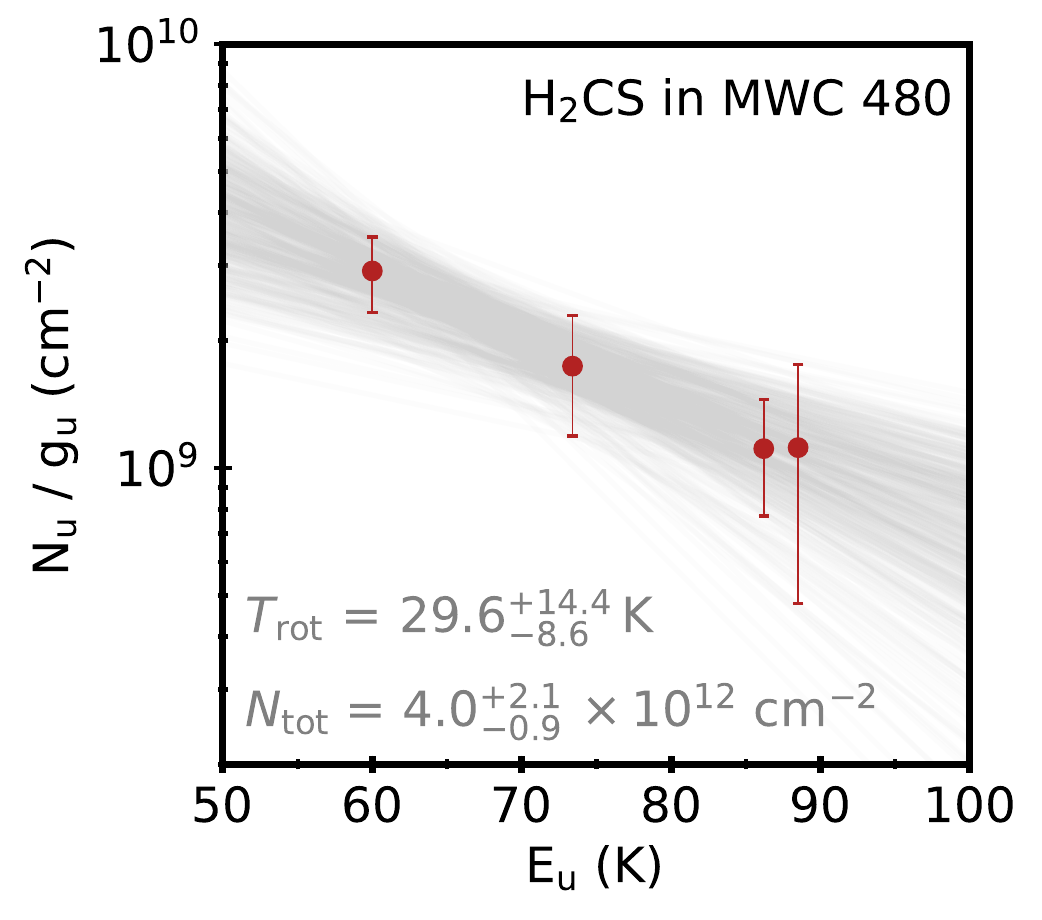}\\
    \caption{Rotational diagrams 
    of {\it(i)} the CS $2-1$, $5-4$, and $6-5$ rotational transitions (top panel), {\it(ii)} the C$^{34}$S $5-4$ and $6-5$ rotational transitions (middle panel), {\it(ii)} \ce{H2CS} $7_{16}-6_{15}$, $8_{17}-7_{16}$, $9_{19}-8_{18}$, and $9_{18}-8_{17}$ rotational transitions (bottom panel), 
    integrated over the outer radius of the molecular line emission, $R_{\mathrm{max}}$, 
    toward MWC~480. A 10\% calibration uncertainty on the flux values is also included.
\label{fig:rot-dia}}
\end{figure}

Using the method described in Section~\ref{subsec:multi_CS_analysis}, we derived the disk-integrated column densities of CS, C$^{34}$S, and \ce{H2CS} in MWC~480. The random draws from the posterior distributions for each molecule are depicted in gray in Fig.~\ref{fig:rot-dia}. The uncertainties are derived from the median and 16th--84th percentiles of the posterior distributions, respectively. The 16th and 84th percentiles are chosen as equivalent to $\pm1 \sigma$ uncertainties on the fit. The results converged toward:
\begin{itemize}
    \item $\tex\simeq22.7^{+2.2}_{-1.8}$~K and $N_{\rm {tot}}\simeq 6.2^{+0.5}_{-0.5} \tdix{12}\cc$ for CS,
    \item $\tex\simeq 14.2^{+16.0}_{-6.8}$~K and $N_{\rm {tot}}\simeq 6.9^{+15.6}_{-2.9} \tdix{11}\cc$ for C$^{34}$S, 
    \item $\tex\simeq 29.6^{+14.4}_{-8.6}$~K and $N_{\rm {tot}}\simeq 4.0^{+2.1}_{-0.9}\tdix{12}\cc$\ for \ce{H2CS}.
\end{itemize} 
This leads to $N$(CS)/$N$(C$^{34}$S) $ \simeq 9^{+20}_{-3}$ and $N$(CS)/$N$(\ce{H2CS})$\simeq 1.6^{+0.8}_{-0.4}$. While the uncertainties on the former do not allow us to draw any firm conclusion, 
the latter is about a factor of two lower than previously found using fewer rotational transitions with a smaller dynamic range in upper energy for \ce{H2CS} \citep[i.e. here we have $E_{\rm{u}}=55.9-88.5$~K versus $E_{\rm{u}}=73.4-88.5$~K in][]{legal2019a}. This illustrates the need for multiple line observations for a given molecule, to better constrain its excitation temperature and column density with rotational diagram methods.

\subsection{Radially resolved column density of CS in MWC~480}

\begin{figure}
    \centering
    \includegraphics[scale=0.75]{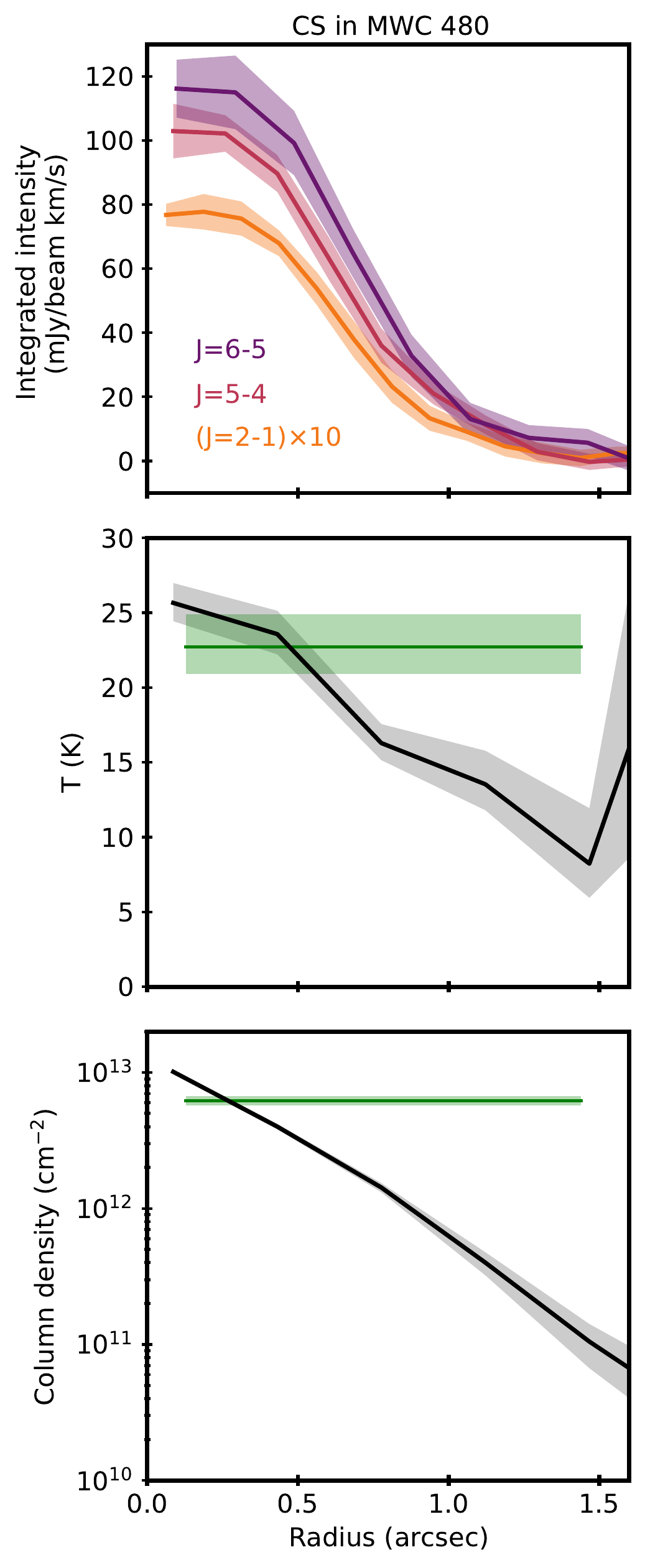}
    \caption{{\it Top panel:} Radially de-projected and azimuthally averaged intensity profiles of the three rotational transitions CS $2-1$, $5-4$, and $6-5$ observed toward the MWC~480 disk. {\it Middle and bottom panels:} Radially de-projected and azimuthally averaged excitation temperature and column density profiles of the MCMC rotational diagram results applied to the aforementioned three CS lines. Median values and uncertainties based on the 16th, 50th, and 84th percentiles of the samples are depicted. For comparison, the disk-integrated CS column density and excitation temperature are over-plotted in green on the middle and bottom panels.}
    \label{fig:CS-radial-rot-dia}
\end{figure}

Applying the same rotational diagram analysis to the radially de-projected and azimuthally averaged intensities, we compute the excitation temperature and column density of CS as a function of the distance from the star. All CS transitions were re-imaged to have matching beam sizes (i.e., $\sim0.5''$). The results are presented in Fig.~\ref{fig:CS-radial-rot-dia}, along with the three CS lines radial intensity profiles. They are in good agreement with the disk-integrated results which are also depicted in Fig.~\ref{fig:CS-radial-rot-dia} to facilitate the comparison. As expected from the CS radial intensity profiles, the CS column density decreases with increasing radius. One can note that the disk average values appear biased toward small distances from the central star because the bulk of the emission is coming from these small distances. So this is why the radially resolved column densities are preferred when derivable. While the temperature gradient is consistent with typical earlier derived radial temperature profiles, it is interesting to notice that the typical model temperatures are higher than the ones derived from the Boltzmann analysis of the observations.

\subsection{disk-integrated column density of CS in MAPS and literature}
\label{subsec:disk-av-CS-in-MAPS}
\begin{figure*}
    \centering
    \includegraphics[scale=1.25]{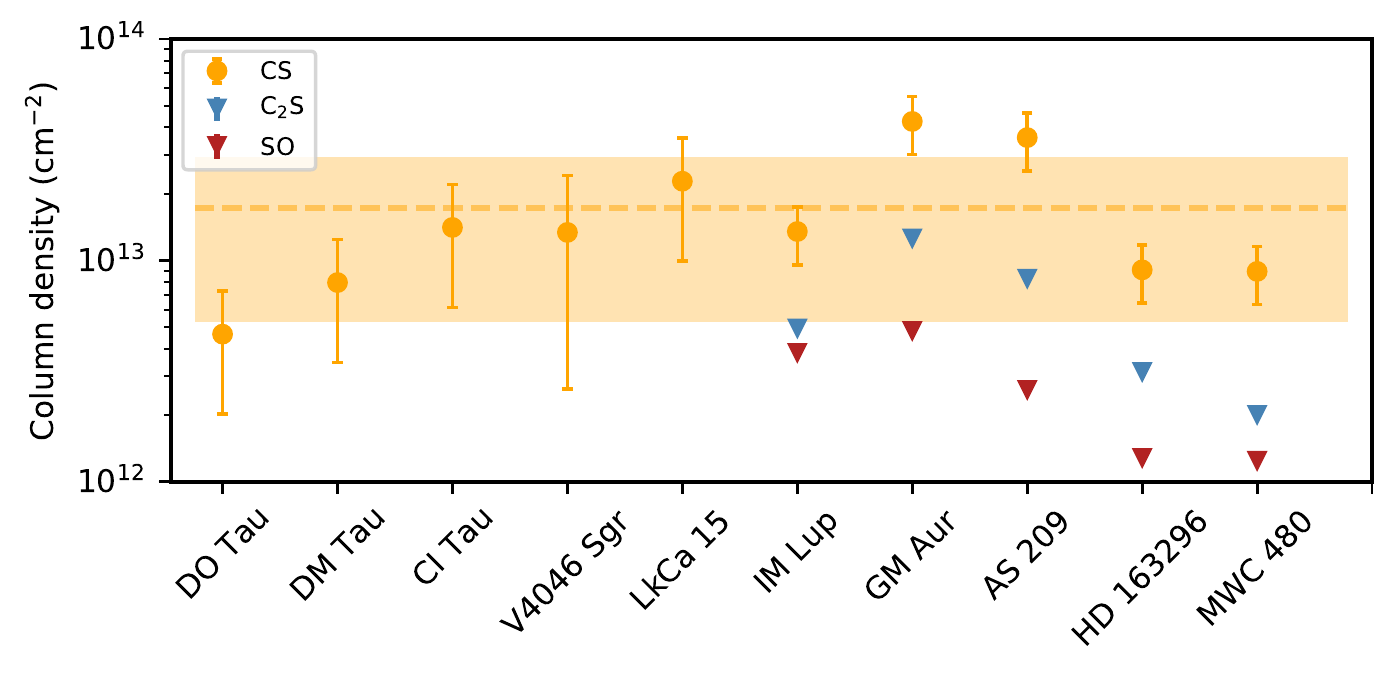}
    \caption{Estimated CS, SO, and \ce{C2S} column densities disk-integrated up to $R_{\mathrm{max}}$ for each MAPS disk (see Table~\ref{tab:obs-list}) and computed for \tex$=10-30$~K. For CS, the disk sample is extended to the additional five disks surveyed in \cite{legal2019a}.
    The disks are sorted by increasing stellar mass. The averaged column density of CS is represented by the dashed orange line and its standard deviation by the orange rectangle. Upper limits are indicated by the downward triangles.}
    \label{fig:CS_SO_CCS_coldens_full_sample}
\end{figure*}

Next, we estimate the disk-integrated CS column densities for the remaining four MAPS disks. As each of these disks only has a single CS transition observed with MAPS (i.e., $2-1$) we fix the excitation temperature to a minimum of 10~K and maximum of 30~K. This temperature range is based on the constraints derived for the MWC 480 disk (see Sect.~\ref{subsec:multi_CS_analysis}), assuming
that CS resides in similar temperature layers in each disk. We calculate the column densities associated with this temperature range using Equations~\ref{eq:Nu} and \ref{eq:boltzmann_dist}.
To enlarge our sample, we extended this calculation to another CS ALMA survey we performed in a sample of five additional disks \citep{legal2019a}. 
The resulting CS disk-integrated column densities are shown in Fig.~\ref{fig:CS_SO_CCS_coldens_full_sample}, sorted by stellar mass. 
The CS disk-integrated column density varies by $\approx 1.5$ order of magnitude across the sample of disks, ranging from $\approx (0.2-6.0)\tdix{13}\cc$. 
There are no obvious trends with stellar mass or spectral type. The two Herbig Ae stars, MWC~480 and HD~163296, are close to the sample average.

\subsection{Upper limits and tentative detections of SO, \ce{C2S}, OCS, and \ce{SO2}}
\label{subsec:tentative-detections}

While the only sulfur-bearing molecular transition targeted in a dedicated SPW within MAPS was $^{12}$CS $2-1$, transitions of \ce{C2S} and SO were covered within the MAPS program (see~Table~\ref{tab:obs-list}). Assuming Keplerian emission and using matched filtering \citep{loomis2018_match_filter} these lines were not detected. To check for non-Keplerian emission, we also imaged these lines using the Briggs weighting with a robust parameter of 1 and a taper of $1''$ to improve the SNR and image quality. The corresponding zeroth moment maps and radially de-projected and azimuthally averaged intensity profiles are shown in Appendix~\ref{app:A}, in Fig.~\ref{fig:mom0maps-radprof-CCS-SO}. 
The integrated intensities and upper limits for the non-detections are reported in Table~\ref{tab:obs-list} for the corresponding $^{12}$CS $2-1$ emitting area. To estimate the upper limits of the SO and \ce{C2S} column densities, we use Eqs.~\ref{eq:Nu} and \ref{eq:boltzmann_dist} with the constraints on the excitation temperature of CS derived in Sect.~\ref{subsec:multi_CS_analysis} (as done in Sect.~\ref{subsec:disk-av-CS-in-MAPS}). The results are over-plotted in red and blue in Fig.~\ref{fig:CS_SO_CCS_coldens_full_sample}.

Finally, two other oxygenated sulfur-bearing molecules, \ce{SO2} and OCS, were observed toward the MWC~480 disk, as part of our complementary ALMA program (see Table~\ref{tab:obs-list}). Figure~\ref{fig:SO2-OCS-mwc480-mom-rad-spec}, in Appendix~\ref{app:A}, shows the zeroth moment maps, radially de-projected and azimuthally averaged intensity profiles, and spectra of \ce{SO2} and OCS respectively. Integrating the intensity over the CS $2-1$ emitting area and FWHM, we find a 3$\sigma$ tentative detection of \ce{SO2} (see Table~\ref{tab:obs-list}) that is also distinguishable from the zeroth moment map where we see a subtle flux peak toward the disk. However, we do not reproduce a detection when using matched filtering (implemented in \texttt{VISIBLE}), nor with velocity shifting and stacking (implemented in \texttt{GoFish}). 
As for OCS, it is not detected and although its radial intensity profile shows a tentative peak toward the central star, the signal shown on the zeroth moment map is shifted from the disk location. Assuming LTE and using the CS excitation temperature derived toward the MWC~480 disk (see Sect.~\ref{subsec:multi_CS_analysis}), we derived upper limits on the column densities that we compare with results from disk chemistry modeling in Fig.~\ref{fig:grid-modeling-coldens}, presented in Sect.~\ref{subsec:C_O_impact}.

\section{Astrochemical modeling}
\label{sec:modeling}

To further investigate the S-chemistry in protoplanetary disks, we computed a grid of astrochemical models tuned to the MWC~480 disk, which is the disk in which we observed the most S-bearing molecules (see Sect.~\ref{subsec:description_obs_C6data}). 

\subsection{Protoplanetary disk physical structure} 

Our fiducial protoplanetary disk astrochemical model is based on the MWC~480 disk model developed in \cite{legal2019a}. It consists of a 2D parametric physical structure in which the chemistry is post-processed (see Sect.~\ref{subsec:disk_model_chem}). We consider here a simplistic physical structure in the sense that the disk is assumed to be symmetric azimuthally and with respect to the midplane. Such a disk physical structure can thus be described in cylindrical coordinates centered on the star along two perpendicular axes characterizing the radius and height in the disk. Figure~\ref{fig:T-n-profiles_and_CS_H2CS} shows the profiles of the gas temperature and density throughout the disk, for which the physical parameters used to compute the physical structure of MWC~480 are summarized in Table~\ref{tab:physical-strcuture-disk-model} and the parameterization is briefly summarized below, following \cite{legal2019a}.

For a given radius $r$ from the central star, the vertical temperature profile is computed following the formalism developed by \cite{dartois2003}:

\begin{equation}
\small
T(z) = \left\{
    \begin{array}{ll}
        T_{\rm{mid}}+(T_{\rm{atm}}-T_{\rm{mid}})\left[ \sin \left(\frac{\pi z}{2z_q}\right) \right]^{2\delta}&\mbox{if} \, z<z_q\\
        T_{\rm{atm}}&\mbox{if} \, z\ge z_q,
    \end{array}
\right.
\label{eq:ture}
\end{equation}
where $T_{\mathrm{mid}}$ and $T_{\mathrm{atm}}$ are respectively the midplane and atmosphere temperatures that
vary as power law of the radii \citep{beckwith1990,pietu2007,legal2019a}. $z_q=4H$ with $H$ the pressure scale height that, assuming vertical hydrostatic equilibrium, can be expressed as follows:

\begin{equation}
H=\sqrt \frac{k_{\rm{B}} \, T_{\rm{mid}} \,r^3}{\mu \,m_{\rm{H}}\, G \,M_\star},
\end{equation}
with $k_{\rm{B}}$ the Boltzmann constant, $\mu=2.4$ the reduced mass of the gas, $m_{\rm{H}}$ the proton mass, $G$ the gravitational constant, and $M_\star$ the mass of the central star. The midplane temperature $T_{\mathrm{mid}}$ is estimated following a simple irradiated passive flared disk approximation \citep[e.g.,][]{chiang1997,dullemond2001}:
\begin{equation}
    T_{\rm{mid}}(r)\approx \left(\frac{\varphi L_\star}{8\pi r^2 \sigma_{\rm{SB}}}\right)^{1/4},
    \label{eq:Tmid_Rc}
\end{equation}
with $L_\star=24~L_\odot$ the stellar luminosity \citep{andrews2013}, $\sigma_{\rm{SB}}$ the Stefan-Boltzman constant and $\varphi=0.05$, a typical flaring angle value \citep[e.g.,][]{brauer2008,baillie2014}. The atmosphere temperature, $T_{\mathrm{atm}}$, is based on observational constraints. So here we consider $T_{\mathrm{atm}}=T_{\mathrm{atm},100\,\rm{au}}(\frac{r}{100\,\rm{au}})$, with $T_{\mathrm{atm},100\,\rm{au}}$=48~K from \cite{guilloteau2011}.

The disk is assumed to be in hydrostatic equilibrium. Thus, for a given vertical temperature profile, the vertical density structure is determined by solving the equation of hydrostatic equilibrium, as described from Eq. (17) to (20) in \cite{legal2019a}.

The surface density of the disk is assumed to follow a simple power law varying as $r^{-3/2}$ \citep{shakura1973,hersant2009}: 
\begin{equation}
\Sigma (r) = \Sigma_{R_c} \left( \frac{r}{R_c}\right)^{-3/2},
\end{equation}
where $\Sigma_{R_c}$ is the surface density at the characteristic radius that can be expressed as function of the mass of the disk, $M_{\rm{disk}}$, and its outer radius, $R_{\rm{out}}$:
\begin{equation}
\Sigma_{R_c}=\frac{M_{\rm{disk}} R_c^{-3/2}}{4\pi \sqrt{R_{\rm{out}}}},
\end{equation}
with here $M_{\rm{disk}}=0.18~M_\odot$ \citep{guilloteau2011}.

The visual extinction profile is derived from the hydrostatic density profile using the gas-to-extinction ratio of $N_{\ce{H}}/A_{\rm{V}}=1.6\times10^{21}$ \citep{wagenblast1989}, with $N_{\ce{H}}=N(\ce{H})+2N(\ce{H2})$ the vertical hydrogen column density of hydrogen nuclei. This gas-to-extinction ratio assumes a typical mean grain radius size of 0.1~$\mu$m and dust-to-mass ratio of 0.01. While the use of a grain size distribution including both large and small grains would be more realistic, its impact on the chemistry remains poorly constrained and would require a dedicated study such as, e.g., the one recently performed in \cite{gavino2021}. We therefore opt for the simpler approximation, which should be sufficient to provide an interpretative framework for the presented observations.

Finally, to compute the UV flux profile we multiplied the UV flux factor impinging on the disk with $e^{-x}$, where $x$ contains the visual extinction profile. The unattenuated UV flux factor, $f_{\rm{UV}}$, at a given radius $r$ depends on both the photons coming directly from the central embedded star and on the photons that are downward-scattered by small grains in the upper atmosphere of the disk. Thus, following \cite{wakelam2016}, we consider:
\begin{equation}
f_{\rm{UV}}=\frac{f_{\rm{UV},R_c}/2}{\left(\frac{r}{R_c}\right)^2+\left(\frac{4\rm{H}}{R_c}\right)^2}.    
\end{equation}

\begin{table}
\begin{center}
\caption{Physical parameters used for our disk models  \label{tab:physical-strcuture-disk-model}}
\begin{tabular}{lc}
\hline\hline
Parameters&MWC~480$^a$\\
\hline
\hline
Stellar mass: $M_\star$ ($M_\odot$) &1.8\\
Disk mass: $M_{\rm{d}}$ ($M_\odot$)& 0.18\\
Characteristic radius: $R_{\rm{c}}$ (au) &100\\
Outer cut-off radius: $R_{\rm{out}}$ (au) &500\\
Density power-law index: $\gamma$&1.5\\
Midplane temperature at $R_{\rm{c}}$$^b$: $T_{\rm{100au}}$(K) &30\\
Atmosphere temperature at $R_{\rm{c}}$: $T_{\rm{100au}}$(K) &48\\
Surface density at $R_{\rm{c}}$: &5.7\\
Temperature power-law index: $q$ &0.5\\
Vertical temperature gradient index: $\beta$&2\\
UV Flux: $f_{{\rm UV},R_c}$ (in \cite{draine1978}'s units) &8500$^c$\\
\hline
\hline
\end{tabular}
\begin{list}{}{}
\item $^a$ These are the values used for the model developed in \cite{legal2019a} and that we are using here to interpret the observations presented in the present work.
\item $^b$ The midplane temperature is estimated from Eq.(~\ref{eq:Tmid_Rc}), the luminosity and a typical flaring angle $\varphi=0.05$.
\item $^c$ from \cite{dutrey2011}, originally computed from the \cite{kurucz1993} ATLAS9 of stellar spectra.
\end{list}
\end{center}
\end{table}

\subsection{Protoplanetary disk chemical model} 
\label{subsec:disk_model_chem}

The disk chemistry is computed time-dependently in 1+1D based on the gas-grain astrochemical code \texttt{Nautilus}, which includes gas-phase, grain-surface, and grain-bulk chemistry \citep{wakelam2017,legal2019a, legal2019b}. This rate-equation gas-grain chemical code follows the formalism described in \cite{hasegawa1992} and \cite{hasegawa1993}. We used the same chemical network as \cite{legal2019a}, which is based on the KInetic Database for Astrochemistry (KIDA)\footnote{(http://kida.obs.u-bordeaux1.fr/)}, and includes recent updates \citep{vidal2017,legal2017,fuente2017,legal2019a}. It contains 589 gas-phase species and 540 solid-state species interacting together through a total of 13402 reactions. Chemical exchanges in between the gas-phase, grain-surface, and grain-bulk phases are included, with adsorption and desorption processes linking the gas and surface phases, and swapping processes linking the mantle and surface of grains. Several desorption mechanisms are taken into account: thermal desorption \citep{hasegawa1992}, cosmic-ray induced desorption \citep{hasegawa1993}, photodesorption and chemical desorption \citep[e.g.,][and references therein]{legal2017}.
In the gas phase typical bi-molecular ion-neutral and neutral-neutral reactions are considered, as well as cosmic-ray induced processes, photoionizations and photodissociations caused by both stellar and interstellar UV photons.

First, we model the chemical evolution of a representative starless dense molecular cloud, with a characteristic age of 1~Myr \citep[e.g.,][]{emelgreen2000,hartmann2001} and typical constant physical conditions: grain and gas temperatures of 10~K, a total gas density of $2\times 10^{4} \ccc$, $\zeta=1\times 10^{-17}\s$ per \ce{H2}, and a visual extinction of 30 mag.
For this first stage model, we consider the initial chemical conditions to be close to diffuse gas conditions, i.e. all the  elements are initially in atomic form (see  Table~\ref{tab:elemental_ab}) except hydrogen which is assumed to be already fully molecular. The elements taken into account in our simulation with an ionization potential lower than that of hydrogen (13.6 eV) are thus assumed to be initially singly ionized, see Table~\ref{tab:elemental_ab}. The outcoming chemical gas and ice compositions of this representative parent molecular cloud serve as the initial chemistry for our 1+1D disk model, for which the physical parameters are described in \cite{legal2019a}. Second, we run the chemistry of our 1+1D disk model up to 1~Myr, the typical chemical age of a disk when dust evolution is not included \citep[e.g.,][]{cleeves2015}, which is the case for our disk model. While the disk chemistry has not reached steady state at that time, its evolution is slow enough that the results presented here hold for a disk twice younger or older.

\begin{table}
\centering
\scriptsize
\caption{Initial Elemental Abundances}
\begin{tabular}{lcc}
\hline\hline
Species & $n_i/n_{\text{H}}$    & Reference\\
\hline
H$_2$   & 0.5\\
He      & 9.0$\times$10$^{-2}$              & 1\\
C$^+$   & 1.7$\times$10$^{-4}$              & 2 \\
N       & 6.2$\times$10$^{-5}$              & 2\\  
O       & $3.4\times 10^{-4} - 1.1\times$10$^{-4}$              & 3\\ 
S$^+$   & $8.0\times 10^{-8} - 1.5\times 10^{-5}$   & 4 \\%
Si$^+$  & 8.0$\times$10$^{-9}$              & 5 \\
Fe$^+$  & 3.0$\times$10$^{-9}$              & 5 \\
Na$^+$  & 2.0$\times$10$^{-9}$              & 5 \\
Mg$^+$  & 7.0$\times$10$^{-9}$              & 5 \\
P$^+$   & 2.0$\times$10$^{-10}$             & 5 \\
Cl$^+$  & 1.0$\times$10$^{-9}$              & 5 \\
F$^+$   & 6.7$\times$10$^{-9}$             & 6 \\
\hline
\label{tab:elemental_ab}
\end{tabular}
\tablecomments{
(1) \cite{wakelam2008}; (2) \cite{jenkins2009};
(3) We varied the oxygen elemental abundance in this range to test the impact of the C/O ratio (see \S~\ref{subsec:C_O_impact}).(see Sect.~\ref{subsec:C_O_impact});
(4) We varied the sulfur elemental abundance in this range to test the impact of the S/H ratio (see \S~\ref{subsec:S_impact});
(5) \cite{graedel1982};
(6) \cite{neufeld2015};}
\end{table}

\subsection{Impact of the S/H ratio}
\label{subsec:S_impact}

\begin{figure*}
    \centering
    \includegraphics[scale=0.4]{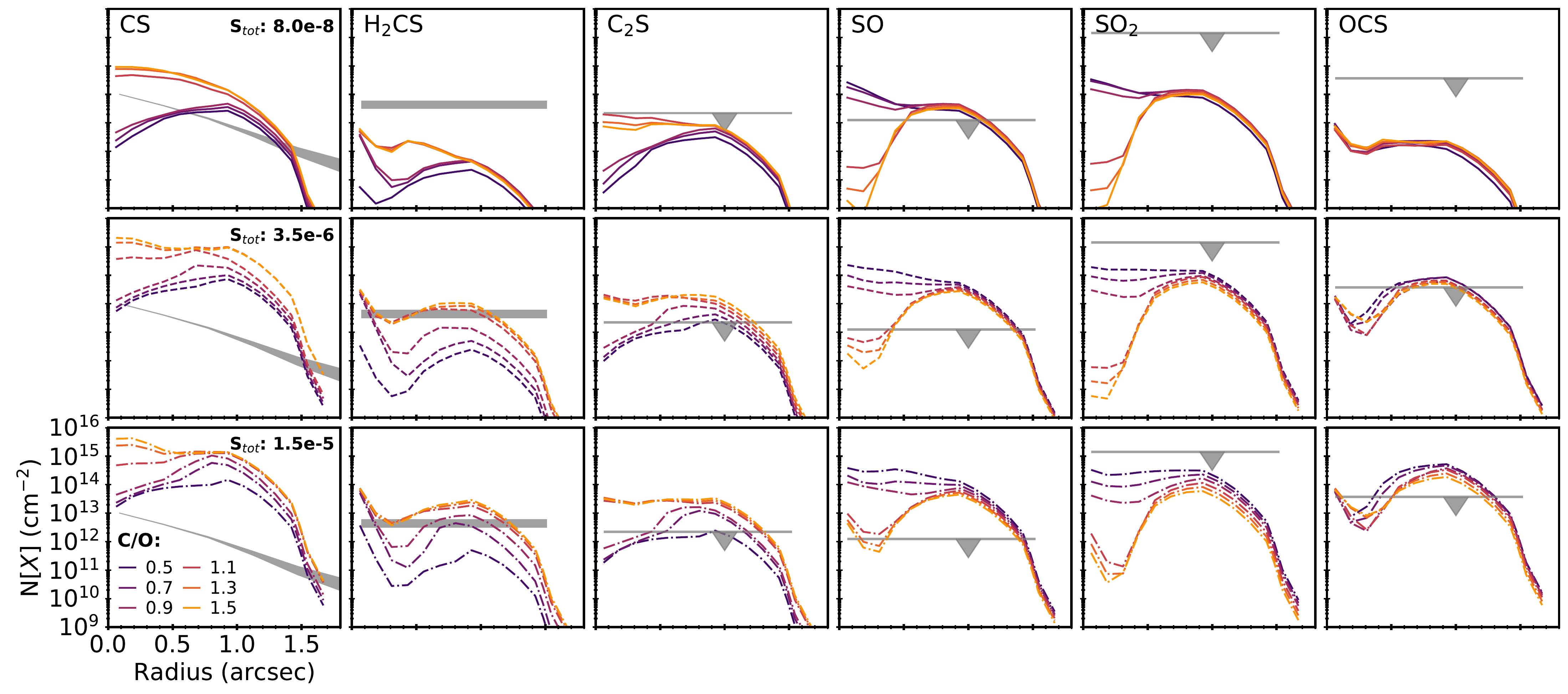}
    \caption{CS, \ce{H2CS}, \ce{C2S}, SO, \ce{SO2}, and OCS modeled column densities tuned to the MWC~480 disk, vertically integrated from the disk upper layer to the midplane and convolved to a resolution of $0\farcs5$ to facilitate the comparison with the observations. The modeled column densities are shown by the solid lines investigating the impact of the C/O and S/H ratios. Observational error bars and upper limits derived toward the MWC~480 disk are indicated in gray. Note that the scales are replicated in all panels.}
    \label{fig:grid-modeling-coldens}
\end{figure*}

In the context of S-bearing molecules, a crucial parameter to study is the S/H elemental ratio, i.e. the total amount of S not locked into refractory compounds and thus available for the volatile S-chemistry. Figure~\ref{fig:grid-modeling-coldens} shows the modeled column densities of CS, \ce{H2CS}, SO, \ce{SO2}, \ce{C2S}, and OCS as function of distance from the central star for a range of C/O ratios (further described in Sect.~\ref{subsec:C_O_impact}) and for three different elemental S/H ratios: the usual highly depleted S-abundance value of $8.0\tdix{-8}$, corresponding to the "low metal" abundances from \cite{graedel1982}, an intermediate S-abundance value of $3.5\tdix{-6}$, corresponding to the value derived in PDR regions \citep{goicoechea2006,legal2019b}, and the solar abundance \citep[$1.5\tdix{-5}$,][]{asplund2009}. For comparison, estimated and upper limits of the column densities of the six S-bearing species we observed toward the MWC~480 disk are also displayed in Fig.~\ref{fig:grid-modeling-coldens}.

While for a low S-elemental abundance, the column density of CS can be reproduced for C/O~$\gtrsim 0.9$, similarly to what has been found to reproduce the column densities of \ce{CH3CN} and \ce{HC3N} in the same disk \citep{legal2019b}, the \ce{H2CS} column densities is under-predicted. On the contrary, no S-depletion, i.e. considering that all the solar S-abundance is available for S-chemistry in disks, allows the reproduction of \ce{H2CS} but cannot reproduce the column density of CS, which is then over-predicted. Models without S-depletion also require different C/O ratios to reproduce the column densities of \ce{C2S} and \ce{SO}. 

Since \ce{H2CS} is a more complex molecule as compared to CS, we suspect that its underproduction in our model is more likely to be due to missing formation pathways than differences in elemental abundances between models and observations. Experimental and theoretical chemical studies are needed to better constrain the formation pathways of \ce{H2CS}.

\begin{figure}
    \centering
    \includegraphics[scale=0.65]{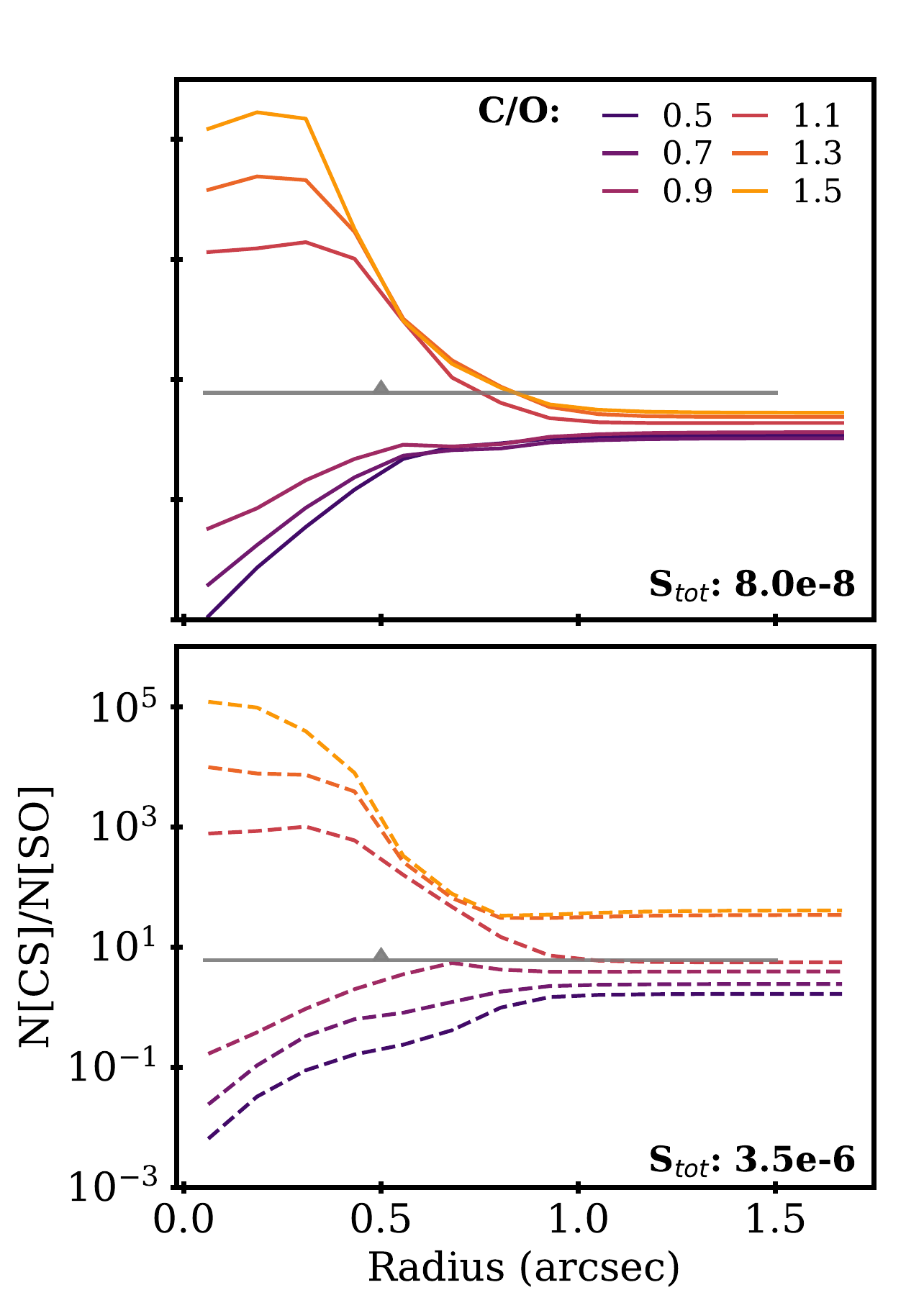}
    \caption{Caculated N(CS)/N(SO) column density ratios for a grid of models tuned to the MWC~480 disk investigating the impact of C/O and S/H ratios. Observations toward the MWC~480 disk} are indicated by the gray horizontal lower limits. Because SO is not detected, the spatial distribution of SO in the disk is unknown. So, we extracted the upper limit on the SO emission for the exact same region of the disk area in which the CS emission is detected (see Section~\ref{subsec:tentative-detections}).
    \label{fig:grid-modeling-coldens_ratios}
\end{figure}

\subsection{Impact of the C/O ratio}
\label{subsec:C_O_impact}

\begin{table*}
\scriptsize
\begin{center}
\caption{Observed versus modeled disk-integrated column densities (in \cc) and CS/SO ratio in MWC~480 out to $1.5''$. \label{tab:obs-vs-model-grid}}
 \renewcommand{\arraystretch}{1.2}
\begin{tabular}{r|cccccccc}
\hline\hline
 \hspace{3cm} & CS & \ce{H2CS} & \ce{C2S} & SO & \ce{SO2} & OCS & CS/SO$^{\dag}$& CS/SO$^{\ddag}$\\ 
\hline
Observed value  & $6.6^{+0.5}_{-0.5} \tdix{12}$ &  $4.0^{+2.1}_{-0.9}\tdix{12}$ & $\lesssim 2.2\tdix{12}$ & $\lesssim 1.2\tdix{12}$ & $\lesssim 1.4\tdix{15}$ & $\lesssim 3.7\tdix{13}$ & $\gtrsim 5.5$\\
\hline
S/H=$8\tdix{-8}$ ; C/O=0.5 & 1.1$\tdix{12}$&	8.1$\tdix{09}$&	1.1$\tdix{11}$&	5.3$\tdix{12}$&	9.6$\tdix{12}$&	1.4$\tdix{11}$ & 0.21 & 0.71\\
\hline
C/O=0.7 &1.5$\tdix{12}$&	4.3$\tdix{10}$&	1.7$\tdix{11}$&	4.8$\tdix{12}$&	1.1$\tdix{13}$&	1.9$\tdix{11}$ & 0.31 &
1.5\\
\hline
C/O=0.9 &1.8$\tdix{12}$&	4.9$\tdix{10}$&	2.1$\tdix{11}$&	3.1$\tdix{12}$&	8.2$\tdix{12}$&	1.5$\tdix{11}$ & 0.58 & 
3.2\\
\hline
C/O=1.1 &2.0$\tdix{13}$&	1.1$\tdix{11}$&	8.6$\tdix{11}$&	1.4$\tdix{12}$&	4.5$\tdix{12}$&	1.3$\tdix{11}$ &  14 &
15\\
\hline
C/O=1.3 &3.3$\tdix{13}$&	1.1$\tdix{11}$&	6.1$\tdix{11}$&	1.3$\tdix{12}$&	3.9$\tdix{12}$&	1.7$\tdix{11}$ & 25\\
\hline
C/O=1.5 &3.5$\tdix{13}$&	1.2$\tdix{11}$&	5.4$\tdix{11}$&	1.1$\tdix{12}$&	3.4$\tdix{12}$&	1.9$\tdix{11}$ & 31\\
\hline
S/H=$3.5\tdix{-6}$ ; C/O=0.5 &2.6$\tdix{13}$&	9.1$\tdix{10}$&	9.0$\tdix{11}$&	7.9$\tdix{13}$&	1.1$\tdix{14}$&	3.1$\tdix{13}$ & 0.33 & 0.92\\
\hline
C/O=0.7 & 3.9$\tdix{13}$&	2.0$\tdix{12}$&	1.5$\tdix{12}$&	3.9$\tdix{13}$&	6.4$\tdix{13}$&	3.0$\tdix{13}$ & 1.0 & 1.1\\
\hline
C/O=0.9 & 7.7$\tdix{13}$&	2.7$\tdix{12}$&	3.0$\tdix{12}$&	2.1$\tdix{13}$&	3.5$\tdix{13}$&	1.9$\tdix{13}$ & 3.7 & 
1.4\\
\hline
C/O=1.1 & 3.1$\tdix{14}$&	5.3$\tdix{12}$&	1.0$\tdix{13}$&	1.0$\tdix{13}$&	2.5$\tdix{13}$&	2.2$\tdix{13}$ & 31 & 11\\
\hline
C/O=1.3 & 7.0$\tdix{14}$&	6.0$\tdix{12}$&	9.6$\tdix{12}$&	9.0$\tdix{12}$&	2.0$\tdix{13}$&	2.0$\tdix{13}$ & 78\\
\hline
C/O=1.5 & 8.2$\tdix{14}$&	6.8$\tdix{12}$&	1.1$\tdix{13}$&	8.1$\tdix{12}$&	1.6$\tdix{13}$&	1.8$\tdix{13}$ & 101\\
\hline
S/H=$1.5\tdix{-5}$ ; C/O=0.5 & 5.8$\tdix{13}$&	4.2$\tdix{11}$&	9.2$\tdix{11}$&	1.7$\tdix{14}$&	2.0$\tdix{14}$&	1.8$\tdix{14}$ & 0.34\\
\hline
C/O=0.7 &1.7$\tdix{14}$&	5.3$\tdix{12}$&	3.3$\tdix{12}$&	8.3$\tdix{13}$&	1.0$\tdix{14}$&	1.5$\tdix{14}$& 2.0\\
\hline
C/O=0.9 & 3.0$\tdix{14}$&	7.8$\tdix{12}$&	5.3$\tdix{12}$&	4.9$\tdix{13}$&	5.5$\tdix{13}$&	1.0$\tdix{14}$& 6.1\\
\hline
C/O=1.1 & 6.2$\tdix{14}$&	1.3$\tdix{13}$&	1.7$\tdix{13}$&	1.8$\tdix{13}$&	3.2$\tdix{13}$&	9.7$\tdix{13}$ & 34\\
\hline
C/O=1.3 & 1.1$\tdix{15}$&	1.5$\tdix{13}$&	1.9$\tdix{13}$&	1.6$\tdix{13}$&	2.3$\tdix{13}$&	7.7$\tdix{13}$ & 69\\
\hline
C/O=1.5 & 1.5$\tdix{15}$&	1.6$\tdix{13}$&	2.0$\tdix{13}$&	1.4$\tdix{13}$&	1.7$\tdix{13}$&	6.1$\tdix{13}$ & 107\\
\hline
\hline
\end{tabular}
\tablenotetext{\dag}{Models w/o dust settling.}
\tablenotetext{\ddag}{Models with dust settling (see Sect.~\ref{subsec:dust-evolution}).}
\end{center}
\end{table*}

The relative gas-phase abundances of the chemical elements are known to strongly influence the chemistry of star-forming regions \citep{vandishoeck1998}. At the onset of star formation, substantial amount of the total budget of the main chemical elements, such as oxygen (O) and carbon (C), are locked in refractory materials. Furthermore, for some of them, huge uncertainties remain on the nature and the form of substantial part of their reservoir. This is in particular the case for oxygen where $\sim40$\% of the O budget remains unaccounted for \citep{whittet2010,jones2019,oberg_bergin_2021} which results in a non-negligible uncertainty on the C/O ratio in the gas-phase.

In order to mimic the differential depletion of volatiles, we varied the C/O ratio from 0.5 to 1.5 (see Table~\ref{tab:elemental_ab}). 
The impact of the gas-phase C/O ratio on the column densities of CS, \ce{H2CS}, SO, \ce{SO2}, \ce{C2S}, and OCS is shown in Fig.~\ref{fig:grid-modeling-coldens} and summarized in Table~\ref{tab:obs-vs-model-grid}. As can be expected, for the carbonated sulfur molecules, i.e. the S-bearing species containing C-S bond, an O-poor chemistry (i.e., a high C/O ratio) results in higher column densities, while the reverse is seen for the oxygenated sulfur molecules (i.e. the molecule containing an O-S bond). This behavior is most prominent in the inner 1$\farcs$0 (i.e., $\sim 160$ au) of the disk for most molecules. Interestingly, the best model to fit the CS data is the most depleted S/H model.

The MAPS observations provide upper limits on SO, which allows us to calculate lower limits on the CS/SO ratio to which we can compare our model.
Figure~\ref{fig:grid-modeling-coldens_ratios} shows how the modeled radial profile of the CS/SO column density ratio varies as a function of the elemental C/O ratio and total amount of sulfur. 
We only consider the two depleted S-abundance models, since we ruled out models with solar S in Section \ref{subsec:S_impact}, based on comparisons between observed and modeled CS radial profiles.
We find that the CS/SO ratio is highly sensitive to the C/O ratio; a change in C/O from 0.5 to 1.5 increases the CS/SO ratio by up to 4 orders of magnitude. This is consistent with previous disk modeling results from \citep[][]{semenov2018}, as well as with cloud chemistry predictions \citep[e.g.][]{bergin1997,nilson2000}. 
We can compare these model results with our observational lower limit of $>5.5$ (see Table~\ref{tab:obs-vs-model-grid}). 
Based on the visual comparison between models and data in Fig.~\ref{fig:grid-modeling-coldens_ratios}, the C/O ratio needs to be $\gtrsim 0.9$ in order to reproduce the CS/SO ratio observation in the MWC~480 disk. We also provide disk-integrated CS/SO ratios for the relevant disk models in Table~\ref{tab:obs-vs-model-grid}, and these confirm that only models with C/O~$>0.9$ are consistent with observations.

\subsection{Impact of dust evolution}
\label{subsec:dust-evolution}

\begin{figure}
    \centering
    \includegraphics[scale=0.65]{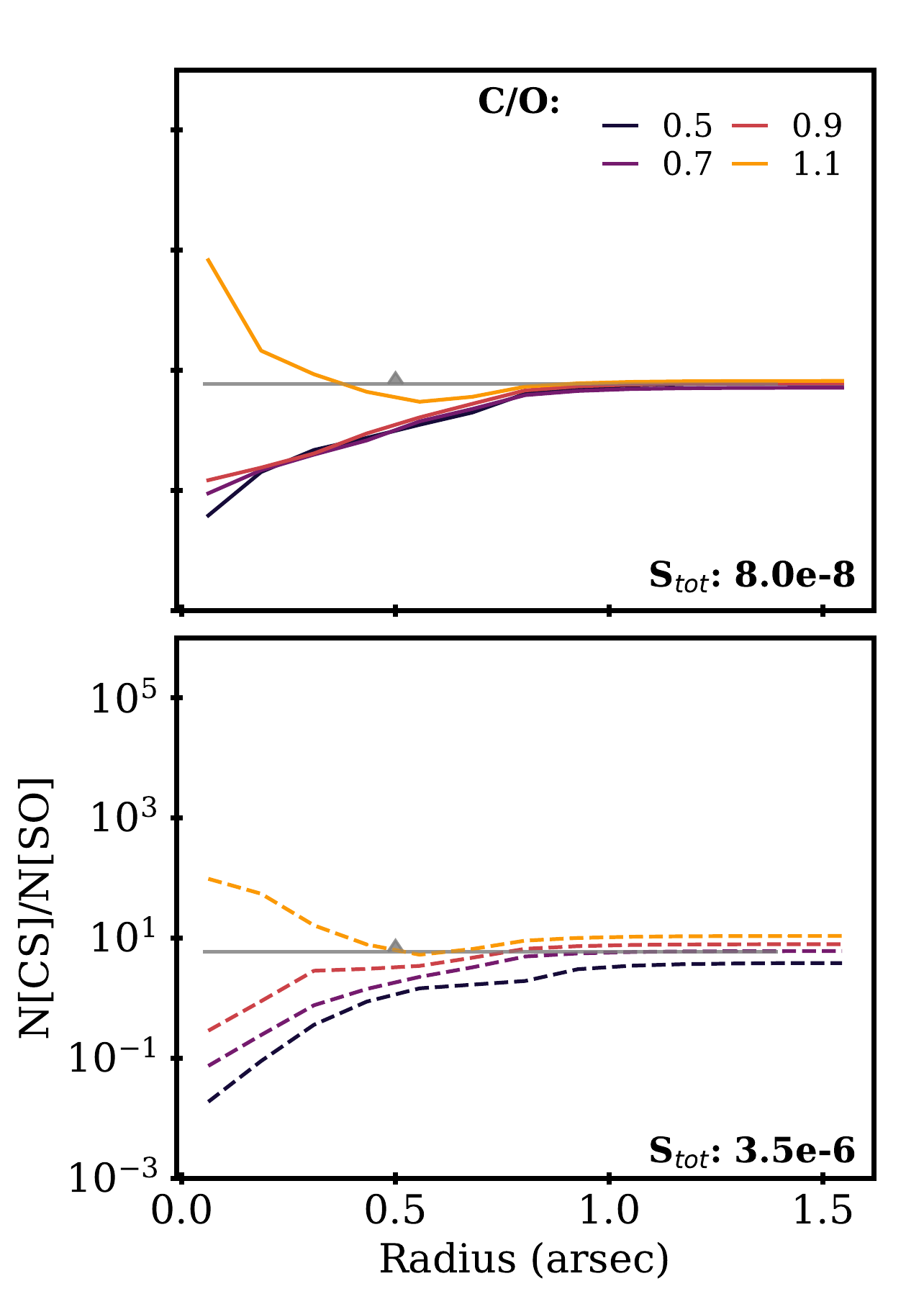}
    \caption{Same as Fig.~\ref{fig:grid-modeling-coldens_ratios} but with models considering dust settling as described in Section~\ref{subsec:dust-evolution}.}
    \label{fig:grid-modeling-coldens_ratios_dust_settling}
\end{figure}

Dust evolution and in particular dust settling can have a profound effect on disk chemistry, including on the ratios of molecules that have been proposed as tracers of C/O. \cite{wakelam2019} recently explored the impact of several disk parameters on the vertically-integrated column densities of a set of molecules observed in the DM~Tau disk. They found that dust settling can have a strong impact on the disk chemistry and can, in particular, enhance the chemical abundances of several carbon-bearing molecules such as \ce{CH3CN} and \ce{HC3N}. To test if dust settling could change our conclusions on the C/O ratio in the MWC 480 disk, we ran an additional set of models, for a smaller grid of C/O values (from 0.5 to 1.1) and including similar dust settling as the fiducial one proposed in \cite{wakelam2019} (i.e. their E2 model, where the settling occurs at $z/h=1$). The results are depicted in Fig.\ref{fig:grid-modeling-coldens_ratios_dust_settling}.

Comparing Figs.~\ref{fig:grid-modeling-coldens_ratios} and \ref{fig:grid-modeling-coldens_ratios_dust_settling}, we see that dust settling indeed influences the variation of CS/SO ratio as function of C/O ratio. With dust settling, the models with varying C/O produce column density ratios within 1 order of magnitude for radius $\gtrsim0.3"$. While without dust settling, at least for the inner disk (i.e. radius $<0.5"$), the models show a spread of 4 to 8 orders of magnitude. Therefore, it seems that if dust settling is present in a disk, one can only derive whether C/O is smaller or larger than 1. Furthermore, one can note that for the outer disk ($>0.7"$), the models with varying C/O are almost indistinguishable when dust settling is present (in particular for the most depleted S/H model).

In summary, despite the apparent dust settling impact on the CS/SO ratio found with our modeling, we find that the results can be consistent with a high C/O ratio under the specific conditions assumed in the modeling.
While out of the scope of the present study, a larger and dedicated deeper study that simultaneously explores the impact of dust settling and different C/O ratios on global disk chemistry would be very interesting to pursue in the future.

\section{Discussion} \label{sec:discussion}

\subsection{Is the CS/SO column density ratio a good C/O ratio proxy?}

As shown in Fig.~\ref{fig:grid-modeling-coldens_ratios}, our modeling results suggest that the CS/SO column density ratio is a promising probe of the C/O ratio in disks (see Section~\ref{subsec:C_O_impact}). We find that an elevated C/O ratio (i.e. a super-solar C/O) is required for the MWC~480 disk in order to reproduce the observed CS/SO ratio. 
A C/O ratio $\gtrsim 0.9$ seems reasonable for the MWC~480 chemistry as it results in both a detectable column density of the S-organic compounds \ce{H2CS} and a good match to the observed column densities of nitriles \citep{legal2019b}. Moreover,while dust settling seems to impact the CS/SO ratio with varying C/O, we are still finding results converging toward a super-solar C/O. It is worth noticing that this is in very good agreement with the results found from other molecules probed within the MAPS program, i.e., CO isotopologues and \ce{C2H}. Using an independent disk model, we found that a super-solar C/O is also required to reproduce the CO isotopologues and \ce{C2H} observations in the same disk \citep{bosman21_CtoO,alarcon2021}. Furthermore, we also checked the predicted water vapor abundances from our models, and in the elevated C/O case they are consistent with the upper limits provided by the WISH project \citep{wish2021}, while water is over-predicted in the low C/O models. Thus, 1) there seems to be a robust convergence toward an elevated C/O ratio and 2) the CS/SO ratio appears to be an additional and independent good probe of the C/O ratio.

\begin{figure}[!h]
    \centering
    \includegraphics[scale=1.1]{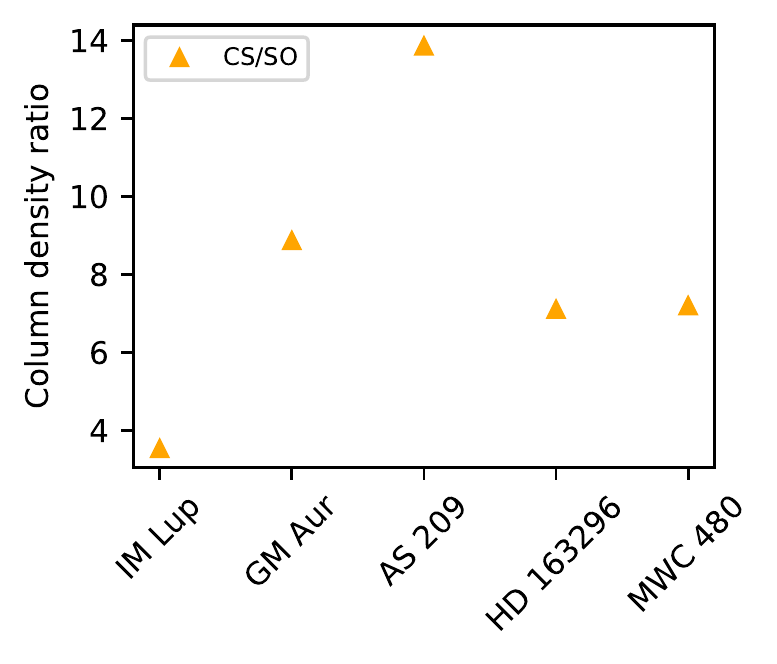}
    \caption{N(CS)/N(SO) column density ratio derived from the MAPS observations.}
    \label{fig:CS_SO_and_CS_CCS_obs_ratios_MAPS}
\end{figure}

In Figure \ref{fig:CS_SO_and_CS_CCS_obs_ratios_MAPS}, we show the lower limits found for the N(CS)/N(SO) column density ratio derived in each of the protoplanetary disks observed with MAPS. Among this sample, the ratio varies by a factor $\sim 10$, which leads to similar C/O ratio constraints for each of the disks, i.e. a super-solar C/O if we consider that our disk model results holds for the other four MAPS disks. However, these preliminary results would need to be corroborated by deeper upper limits on SO and further modeling investigations which would be address in forthcoming studies. Additional CS/SO measurements toward a larger sample of protoplanetary disks would also be a good way to measure how common is the C/O ratio expected to be elevated in disks. 

\subsection{Interpretation of disk S/H ratio}

In protoplanetary disks, the S/H elemental ratio has been much less studied, and therefore less well constrained, than the C/O and C/H ratios. As of today, we still do not know what the major S reservoir(s) in disks is (are) and in which form it resides (solid or gaseous). However, this is an important parameter to constrain as well, not only to solve the current disk modeling tension found to interpret the high \ce{H2CS}/CS ratio in MWC~480, but also because, more generally, many S-bearing species are observed in comets and do play an important role in the building-up of pre-biotic molecules and on planet habitability.

Recently, based on the abundances of B star photospheres, \cite{kama2019} found that $\sim81-97\%$ of the S-budget should be locked in disk refractory material. Following the findings of \cite{keller2002}, the former authors proposed that most of the sulfur should be locked in the form of solid FeS in disks, rather than in polymeric S$_n$ ($n=2-8$) molecules, where the latter has been proposed for decades as a potential hidden S-reservoir in the ISM \citep[e.g.,][]{wakelam2004}. 
However, the observations of solid FeS has not yet been confirmed in disks \citep{keller2002} whereas hints of solid S$_n$ were recently reported on the comet 67P/C-G with the detections of S$_2$, S$_3$, and S$_4$ from the Rosetta mission \citep{calmonte2016}, although these detections constitute only $\sim0.2\%$ of the total detected S-content of 67 P/C-G. 

The comparison with comets' sulfur-bearing molecules content is, however, instructive. In particular, it is interesting to note that in the inventory of the molecular abundances detected in comets compiled by \cite{bockelee-morvan2017}, a \ce{H2CS}/CS ratio of $\approx 0.45$ is reported, i.e. close to the value we measured in the MWC~480 protoplanetary disk \citep[see Section~\ref{subsec:disk-int-col-dens-MWC480} and][]{legal2019a}. Another relevant point stressed in the \cite{bockelee-morvan2017} review is that one of the most abundant S-bearing molecules detected in comets is OCS, which could be another potential S-reservoir. OCS is indeed the only S-bearing molecule unambiguously detected in ice mantles so far \citep{geballe1985,palumbo1995}. Furthermore, astrochemical shock modeling benchmarked to protostellar shock observations predicted that$\gtrsim$ 50\% of the sulfur ice reservoir resides in OCS \citep{podio2014,holdship2016}. The latter is therefore a promising S-reservoir to interpret the high \ce{H2CS}/CS ratio we observed toward MWC~480 that is not reproducible with our model (see Section~\ref{sec:modeling}). 

\subsection{Could the sulfur organic chemistry be underappreciated in models?}

In our model, and as previously described in \cite{legal2019a}, \ce{H2CS} is mainly formed from the following neutral-neutral and electronic dissociative recombination gas-phase reactions:
\begin{equation}
\ce{S + \ce{CH3} -> \ce{H2CS} + H},
\end{equation}
\begin{equation}
\ce{\ce{H3CS+} + \ce{e-} -> \ce{H2CS} + H},
\end{equation}
with \ce{H3CS+} originating from the \ce{S+ + CH4} reaction.

It is also formed, for a smaller contribution, from gas-grain chemistry where \ce{H2CS} is produced from successive hydrogenation on icy dust mantles and released for $\sim 1$\% in the gas phase by chemical reactive desorption. However, in our current model, there is no consideration of OCS grain surface processing that could lead to the formation of S-organics such as \ce{H2CS} and thus maybe help in better reproducing the observations.
Because \ce{H2CS} is a more complex species compared to CS, we suspect that its underproduction is more likely due to missing formation pathways than an un-depleted S gas-phase reservoir. The latter is inconsistent with CS observations (see Section~\ref{subsec:S_impact}).
Laboratory experiments and theoretical chemical calculations for such mechanisms are required to further test our hypothesis. While the S-reservoir could be changing from the inner to the outer disk regions, dedicated disk resolved S-observations are also needed to further investigate the nature and identity of the S-reservoir in disks, and how and if there is any chemical inheritance from molecular cloud stage to the planet-forming environment.

\subsection{CS disk-emission asymmetries}
\label{subsec:CS-asym}

In Sect.~\ref{subsec:CS-morpho}, we highlight asymmetries in the CS $2-1$ emission spatial distribution toward four of the five targeted disks. Intriguingly, these asymmetries are not observed in the other molecular lines targeted within MAPS nor in higher CS transitions \citep[i.e., the $5-4$ and $6-5$ transitions published in][]{legal2019a}. We also do not see any such asymmetries in the dust emission of any of the targeted disks. So these asymmetries appears to be different than, for instance, the one detected in CS $J=7-6$ toward the HD~142527 disk \citep{vanderplas2014}. However, we should note that in the present study we are reporting CS $2-1$ asymmetries which are likely emitting from disk layers closer to the midplane than the CS $7-6$ would. Toward the 
GM~Aur, AS~209, and HD~163296 disks, the CS $2-1$ emission asymmetries seem to correlate with their respective disk inclination, i.e. the CS $2-1$ emission is brightest in the near side of the disk (see~Fig.~\ref{fig:asym-disks}). However, toward MWC~480, we observe the reverse: the CS $2-1$ emission is brightest in the far side of the disk. Here we investigate what could cause such asymmetries.

The HD~163296 disk is known to harbor both a jet \citep{grady2000} and a disk wind. The latter was discovered through $^{12}$CO observations by \cite{klaassen2013} and is also further characterized using CO isotopologue observations as a part of MAPS \citep{booth2021maps}.
According to the geometry of jet, wind and viewing angle proposed in \cite{ellerbroek2014}, the far side of the disk is supposed to be viewed through the disk's wind and jet, which is also the disk side where we found the CS $2-1$ emission to be the weakest.
Thus, a speculative interpretation could be that the disk's wind and jet impact the CS content, or simply the emission of the $2-1$ line, and could therefore explain the decline in CS $2-1$ flux in the side of the disk affected by the wind and jet. For instance, if the wind impacts the local C/O ratio thus it could impact the local disk chemistry and maybe the total amount of the CS bulk lying closer to the midplane layer; or/and if the wind is also made of dust, the line emission below the wind with respect to the observer could be hampered due to dust wind opacity. However, follow-up observations of CS toward both the HD~163296 disk and wind would be required to test this hypothesis. 

Similarly, MWC~480 is known to be driving a bipolar jet aligned with the disk semi-minor axis \citep{grady2010}. Notably, the jet flow appears denser in the SW direction which could explain the decrease in CS $2-1$ emission we observe in the same direction. A better characterization of this jet is required to assess how it could impact the CS $2-1$ disk emission.

Hints for a disk wind are also found toward the AS~209 disk \citep[e.g.,][and references therein]{banzatti2019,fang2018}, but, to our knowledge, the orientation and spatial distribution of the latter remain to be determined. Another point to mention about the AS~209 disk is that its west half side is known to be cloud-contaminated \citep{oberg2011_DISCS,huang2016,teague2018_as209}. While it strongly impacts the $^{12}$CO $2-1$ and \ce{HCO+} $1-0$ line emission in this disk \citep[see Fig.~2 and 4 in][]{law21_rad}, this cloud-contamination does not match with the CS $2-1$ asymmetries we are finding in the present work.

As for GM~Aur, \cite{macias2018} discuss the possibility of a radio jet. Furthermore, the GM~Aur disk is also the only transitional disk of our sample -- i.e., the only one with a central dust cavity -- and, as characterized by its complex gas structures, it is known to be affected by much more prominent gas dynamics than the other disks of our sample \citep[e.g.,][and reference therein]{huang2021}. Therefore, the CS $2-1$ asymmetry of this disk is probably the least difficult to justify but would require further observations to be linked with the other gas kinematics features observed in this disk.

While all these hypothesis seem appealing, further investigations are required to truly determine the nature of these CS $2-1$ asymmetries and in particular to identify if they are tracing one specific characteristic of disk evolution or if they could be explained by multiple phenomena.

\section{Conclusion} \label{sec:conclusion}

We presented ALMA observations of S-bearing molecules observed toward the five protoplanetary disks targeted by the MAPS ALMA Large Program, orbiting the IM~Lup, GM~Aur, and AS~209 T~Tauri stars, and the two Herbig Ae stars HD~163296 and MWC~480. 

Our main findings are summarized below:
\begin{enumerate}
    \item The CS $2-1$ line was observed within MAPS and detected toward all five disks displaying a variety of radial intensity profiles and spatial distributions across the sample, including intriguing apparent azimuthal asymmetries.
    \item Using complementary ALMA observations of CS $5-4$ and $6-5$ in one of the disks, i.e., the MWC~480 disk, allows us to assess the CS column density across the full sample, assuming a temperature in the range 10-30~K, which results in \ntot(CS)$\approx(0.2-5)\tdix{13}\cc$ .
    \item \ce{C2S} and SO lines were also covered within MAPS. While no detection can be robustly claimed from these observations, we provide upper limits on their column densities, with upper limits in the range $10^{12}-10^{13}$\ccc for \ce{C2S} and $[1-5]\times10^{12}$\ccc for SO. In particular, we used the upper limit on SO to derive lower limits on the CS/SO ratio across the MAPS sample, which is found to range from $\sim 4$ to 14.
    \item Using complementary ALMA programs, we find $N$(\ce{H2CS})/$N$(CS)$\approx 2/3$ in MWC~480. This high ratio suggests that substantial S-reservoirs in disks may be in the form of S-organics (i.e., C$_x$H$_y$S$_z$).
    \item Using astrochemical disk models, we find that the CS/SO ratio is a promising probe for the elemental C/O ratio. CS/SO varies by more than 4 orders of magnitude when C/O varies from 0.5 to 1.5. 
    \item For MWC~480, without considering dust settling, we find C/O $\gtrsim 0.9$, consistent with constraints from nitriles observations \citep{legal2019b}. When considering dust settling, one can only derive whether C/O is smaller or larger than 1, but this remains consistent with a high C/O ratio under the specific conditions assumed in the modeling. More interestingly, this is confirmed with independent disk chemistry models predicting super-solar C/O based on the CO and \ce{C2H} MAPS data \citep{bosman21_CtoO}.
    \item We find a depleted gas-phase S/H ratio, suggesting either that part of the sulfur reservoir is locked in solid phase or that it remains in an unidentified gas-phase reservoir.
    More sulfur observations are required to confirm this and, to a larger extent, to identify the nature of the S-reservoir(s).
    
    Together these results illustrate the importance of sulfur chemistry in protoplanetary disks, demonstrating that, not only, sulfur-bearing molecules observations in such disks can serve to constrain the sulfur chemistry itself and its reservoir(s), but also that sulfur-bearing molecules are powerful tools to constrain other fundamental parameters, such as the elemental C/O ratio. Furthermore, sulfur-bearing molecules seem to uniquely probe disk gas substructures, but this require deeper observations to be further investigated and confirmed. Therefore, to fully comprehend the role of sulfur in disks, further theoretical and observational studies on the sulfur chemistry in disks are still required. 
    \end{enumerate}

\acknowledgments

We thank the referees for their valuable comments which helped us to improve the quality of this manuscript. R.L.G. also thanks Clément Baruteau for helpful discussions.
This paper makes use of the following ALMA data: ADS/JAO.ALMA\#2018.1.01055.L and ADS/JAO.ALMA\#2018.1.01631.S. ALMA is a partnership of ESO (representing its member states), NSF (USA) and NINS (Japan), together with NRC (Canada), MOST and ASIAA (Taiwan), and KASI (Republic of Korea), in cooperation with the Republic of Chile. The Joint ALMA Observatory is operated by ESO, AUI/NRAO and NAOJ. The National Radio Astronomy Observatory is a facility of the National Science Foundation operated under cooperative agreement by Associated Universities, Inc.

R.L.G. acknowledges support from CNES fellowship grant.
K.I.\"O. and R.L.G. acknowledge support from a Simons Foundation award (SCOL \# 321183, KO) and an NSF AAG Grant (\#1907653). R.T. acknowledges support from the Smithsonian Institution as a Submillimeter Array (SMA) Fellow. 
C.J.L. acknowledges funding from the National Science Foundation Graduate Research Fellowship under Grant DGE1745303. C.W.~acknowledges financial support from the University of Leeds, STFC and UKRI (grant numbers ST/R000549/1, ST/T000287/1, MR/T040726/1).
E.A.B. and A.D.B. acknowledge support from NSF AAG Grant \#1907653.
F.M. acknowledges support from ANR of France under contract ANR-16-CE31-0013 (Planet-Forming-Disks) and ANR-15-IDEX-02 (through CDP ``Origins of Life").
S. M. A. and J. H. acknowledge funding support from the National Aeronautics and Space Administration under Grant No. 17-XRP17 2-0012 issued through the Exoplanets Research Program. 
J.H. acknowledges support for this work provided by NASA through the NASA Hubble Fellowship grant \#HST-HF2-51460.001-A awarded by the Space Telescope Science Institute, which is operated by the Association of Universities for Research in Astronomy, Inc., for NASA, under contract NAS5-26555.
Y.A. acknowledges support by NAOJ ALMA Scientific Research Grant Numbers 2019-13B, and Grant-in-Aid for Scientific Research 18H05222 and 20H05847. 
A.S.B. acknowledges the studentship funded by the Science and Technology Facilities Council of the United Kingdom (STFC).
G.C. is supported by NAOJ ALMA Scientific Research Grant Code 2019-13B.
J.B.B. acknowledges support from NASA through the NASA Hubble Fellowship grant \#HST-HF2-51429.001-A, awarded by the Space Telescope Science Institute, which is operated by the Association of Universities for Research in Astronomy, Inc., for NASA, under contract NAS5-26555. 
L.I.C. gratefully acknowledges support from the David and Lucille Packard Foundation and Johnson \& Johnson's WiSTEM2D Program.
IC was supported by NASA through the NASA Hubble Fellowship grant HST-HF2-51405.001-A awarded by the Space Telescope Science Institute, which is operated by the Association of Universities for Research in Astronomy, Inc., for NASA, under contract NAS5-26555. 
V.V.G. acknowledges support from FONDECYT Iniciaci\'on 11180904 and ANID project Basal AFB-170002.
J.D.I. acknowledges support from the Science and Technology Facilities Council of the United Kingdom (STFC) under ST/T000287/1.
H.N. acknowledges support by NAOJ ALMA Scientific Research Grant Code 2018-10B and Grant-in-Aid for Scientific Research 18H05441.
K.R.S. acknowledges the support of NASA through Hubble Fellowship Program grantHST-HF2-51419.001, awarded by the Space Telescope Science Institute,which is operated by the Association of Universities forResearch in Astronomy, Inc., for NASA, under contract NAS5-26555. 
T.T. is supported by JSPS KAKENHI Grant Numbers JP17K14244 and JP20K04017.
Y.Y. is supported by IGPEES, WINGS Program, the University of Tokyo.
K.Z. acknowledges the support of the Office of the Vice Chancellor for Research and Graduate Education at the University of Wisconsin – Madison with funding from the Wisconsin Alumni Research Foundation, and the support of NASA through Hubble Fellowship grant \#HST-HF2-51401.001. awarded by the Space Telescope Science Institute, which is operated by the Association of Universities for Research in Astronomy, Inc., for NASA, under contract NAS5-26555. 

\vspace{2cm}

\facilities{ALMA}

\software  {\texttt{CASA}} \citep[][]{mcmullin2007}, {\texttt{Astropy} \citep{astropy}, \texttt{Matplotlib} \citep{matplotlib}, \texttt{NumPy} \citep{numpy_article},  \texttt{emcee} \citep{emcee2013},  \texttt {SciPy} \citep{scipy}, \texttt{scikit-image} \citep{scikit-image}, \texttt{Gofish}\citep{teague2019},\texttt{VISIBLE}\citep{loomis2018_match_filter}}

\newpage
\appendix

\section{Imaging of SO, \ce{C2S}, OCS, and \ce{SO2}}
\label{app:A}
\begin{figure*}
    \centering
    \includegraphics[scale=0.43]{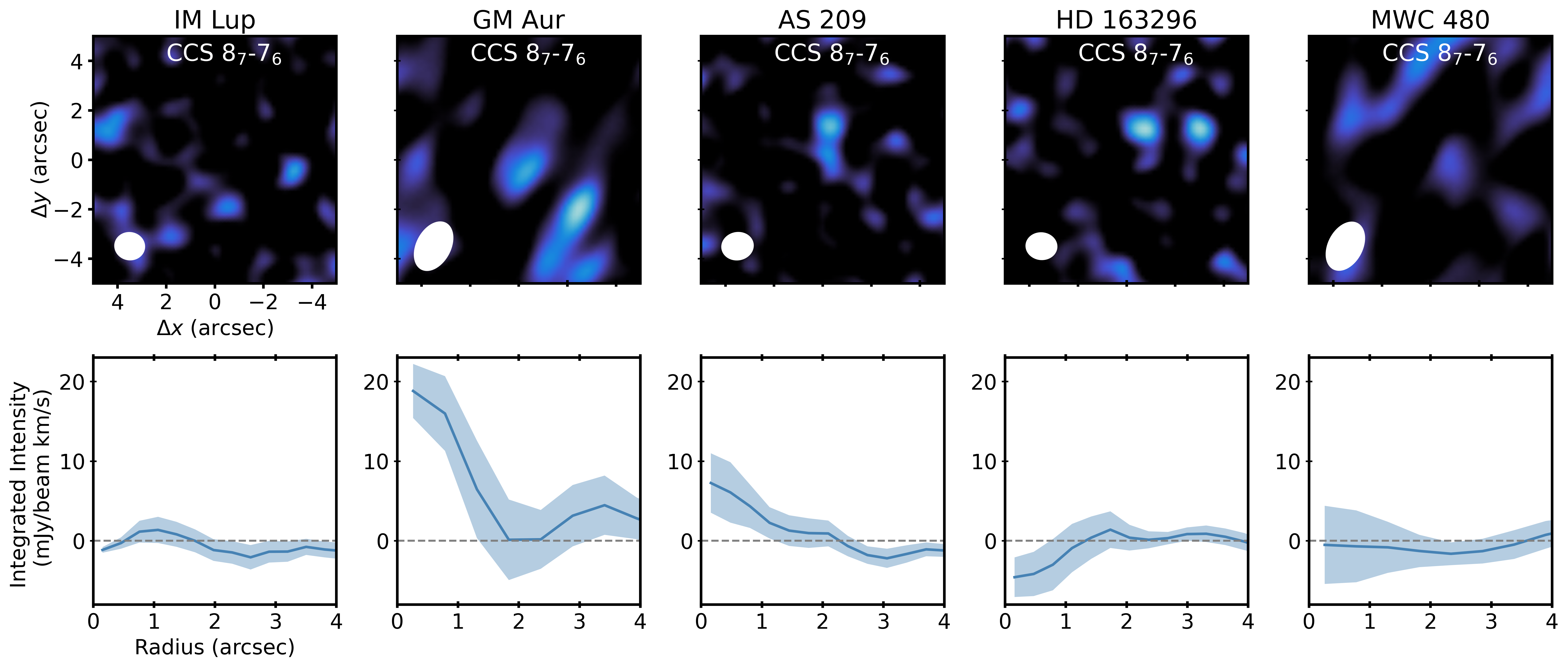}
    \includegraphics[scale=0.43]{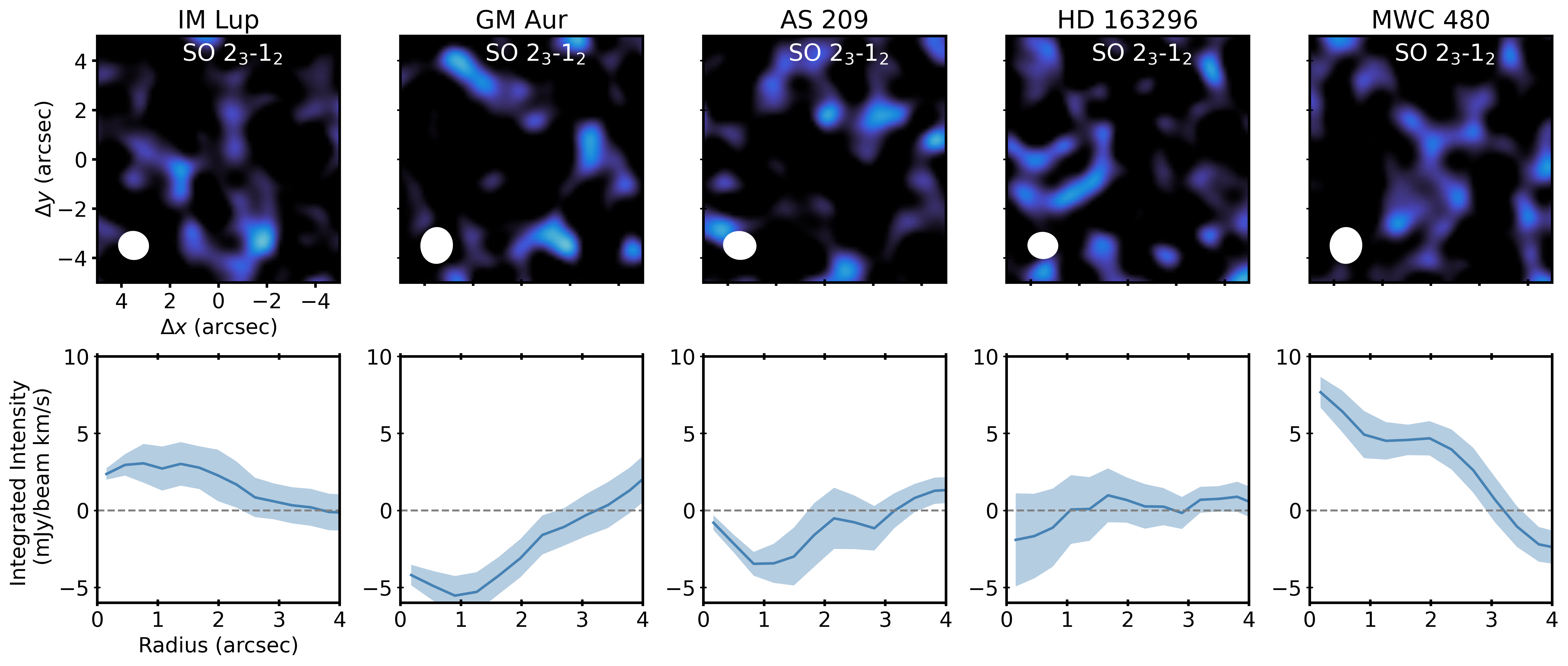}
    \caption{{\it First and third rows:} Zeroth moment maps of
    \ce{C2S} $8_7-7_6$ (upper row) and SO $2_3-1_2$ (third row). Synthesized beams are shown in the
    lower left corner of each panel. 
    {\it Second and fourth rows:} Radially de-projected and azimuthally averaged intensity profiles of the \ce{C2S} $8_7-7_6$ (second row) and SO $2_3-1_2$ (fourth row) emission lines.}
    \label{fig:mom0maps-radprof-CCS-SO}
\end{figure*}

\begin{figure*}
    \centering
    \includegraphics[scale=0.5]{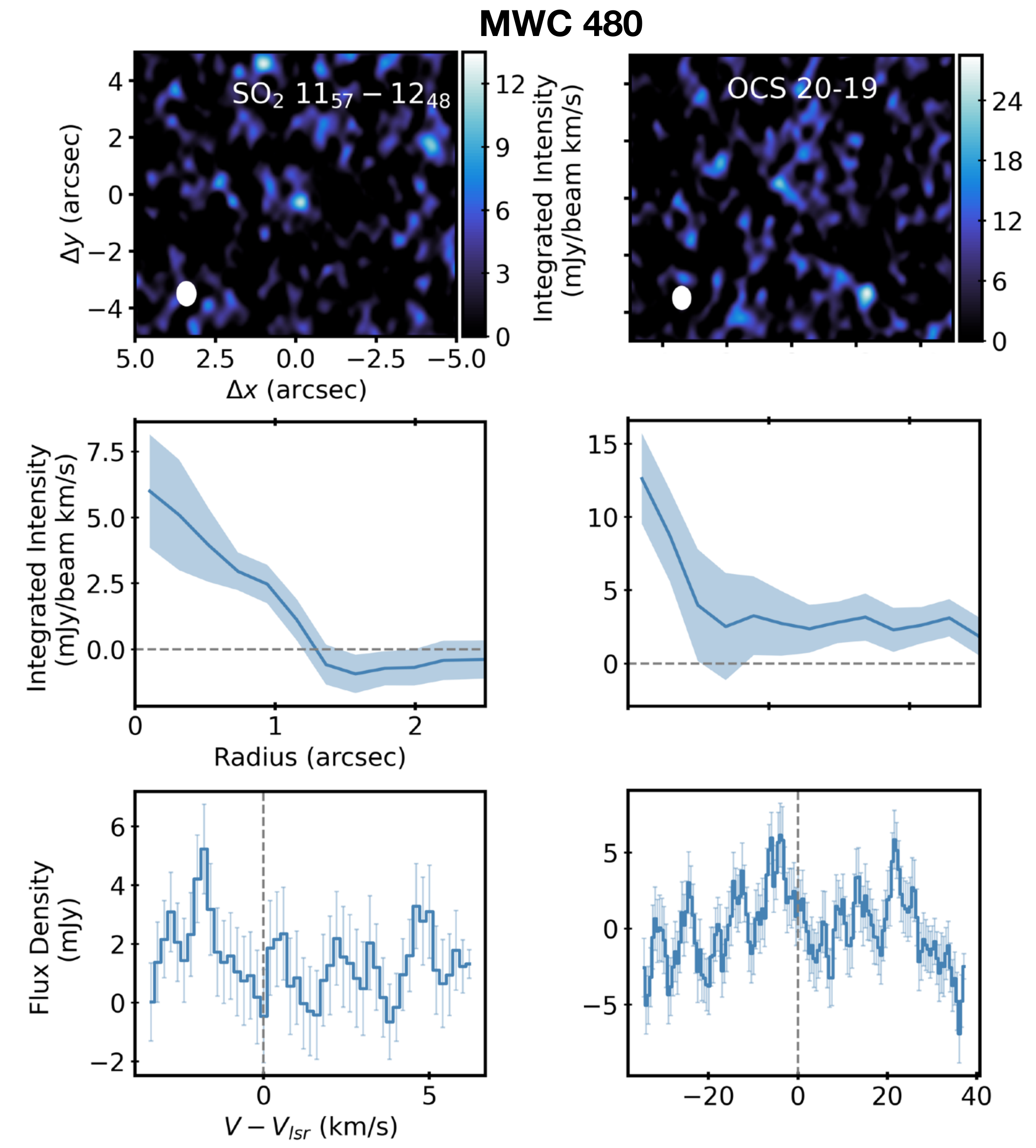}
    \caption{Zeroth moment map (top panels), radially de-projected and azimthally averaged intensity profile within $1\sigma$ - similarly too Fig.~\ref{fig:MAPS-CS21-obs-overview} and \ref{fig:C34S-H2CS-mwc480-mom-rad-spec} (middle panels), and shifted and stacked disk-integrated line spectra within $1\sigma$ (bottom panels). These last  uncertainties are calculated on a per channel basis, taking into account de-correlation along the spectral axis \citep[see also][]{yen2016,ilee2021}.}
    \label{fig:SO2-OCS-mwc480-mom-rad-spec}
\end{figure*}

In this section, we present the zeroth moment map, radially de-projected and azimthally averaged intensity profile, and shifted and stacked disk-integrated spectrum zeroth moment maps, radial intensity profiles for the strongest \ce{C2S} and SO lines serendipitously covered by MAPS (Fig.~~\ref{fig:mom0maps-radprof-CCS-SO}) and for the \ce{SO2} and OCS lines covered in our complementary Cycle 6 ALMA data (Fig.~\ref{fig:SO2-OCS-mwc480-mom-rad-spec}). All of them are derived for the same emitting area as the detected CS $2-1$ transition (see Table~\ref{tab:obs-list}).


\bibliography{biblio}
\bibliographystyle{aasjournal}

\end{document}